\documentclass[preprint]{aastex}

\usepackage{graphicx}
\usepackage{wrapfig}
\usepackage{natbib}

\newcommand{\be}{\begin{equation}}
\newcommand{\ee}{\end{equation}}
\newcommand{\nn}{\mbox{} \nonumber \\ \mbox{} }
\newcommand{\ba}{\begin{eqnarray}}
\newcommand{\ea}{\end{eqnarray}}
\newcommand{\om}{\omega}
\newcommand{\Alfven}{Alfv\'{e}n }

\newcommand{\curl}{{\rm curl\, }}
\newcommand{\A}{{\bf A}}
\newcommand{\E}{{\bf E}}
\newcommand{\B}{{\bf B}}

\newcommand\eg{{\it{e.g.\ }}}

\newcommand{\Bf}{{magnetic field}}
\newcommand{\Bfs}{{magnetic fields}}
\newcommand{\Ef}{{electric  field}}

\newcommand{\NS}{neutron star}
\newcommand{\NSs}{{neutron stars}}
\newcommand{\EM}{electromagnetic}

\newcommand{\ms}{magnetosphere}
\newcommand{\mss}{magnetospheres}

\newcommand{\LC}{light cylinder}
\newcommand{\Lf}{Lorentz factor}

\begin{document}

\title{Escaping of Fast Radio Bursts}
\author{Maxim Lyutikov, \\
Department of Physics and Astronomy, Purdue University, 
 525 Northwestern Avenue,
West Lafayette, IN
47907-2036, USA; lyutikov@purdue.edu}

\begin{abstract}
We reconsider  the  escape of high brightness coherent emission of Fast Radio Bursts (FRBs)  from   magnetars' \mss,  and   conclude that  there are numerous ways for the powerful FRB pulse to avoid nonlinear  absorption.  Sufficiently strong surface magnetic fields,  $\geq 10\%$ of  the quantum   field,  limit the waves' non-linearity to moderate values. For weaker fields, the    \Ef\ experienced  by a particle is limited by a 
combined ponderomotive and parallel-adiabatic  forward  acceleration   of charges by the incoming FRB pulse along the  \Bf\ lines  newly opened   during FRB/Coronal Mass Ejection  (CME). As a result, particles surf the weaker  front   part of the pulse, experiencing low radiative losses, and are cleared from  the \ms\ for the bulk of the pulse to propagate.  We  also find: 
  (i) for propagation across \Bf, the O-mode  suffers much smaller dissipation than the  X-mode; (ii)  quasi-parallel propagation  suffers minimal dissipation;  (iii)  initial mildly relativistic radial plasma flow further reduces losses;
 (iv) for oblique propagation of a pulse with limited transverse size,  the leading part of the pulse would  ponderomotively  sweep the plasma aside.  
   \end{abstract}

\maketitle 

\section{Introduction}

Observations of  correlated radio and X-ray bursts \citep{2020Natur.587...54C,2021NatAs...5..372R,2020Natur.587...59B,2020ApJ...898L..29M,2021NatAs.tmp...54L} 
established the  FRB-magnetar connection. There is a long list of arguments in favor of magnetospheric  {\it loci}  of  FRBs   \citep[\eg][]{2003MNRAS.346..540L,2013arXiv1307.4924P,2016MNRAS.462..941L,2020arXiv200505093L}, as opposed, \eg\ to the wind  \citep[\eg][]{2014MNRAS.442L...9L,2017ApJ...843L..26B,2019MNRAS.485.4091M,2022arXiv220911136T} (see also reconsideration of wind dynamics by \cite{2023MNRAS.524.6024S}, arguing against appearance of shocks). For example, 
temporal coincidence between the radio and X-ray profiles, down to milliseconds  is a strong argument in favor of magnetospheric origin 
\citep{2020arXiv200505093L}: we know that X-ray are magnetospheric events, as demonstrated by the periodic oscillations seen in giant flares  \citep{2005Natur.434.1107P,2005Natur.434.1098H}.

In fact, strong \Bfs\  are needed in the emission region to suppress strong `normal'' (non-coherent)  loses in  magnetar \mss.  In the absence of strong guide-field a coherently emitting particle will lose energy on time scales shorter than the coherent low frequency wave.  \cite{2016MNRAS.462..941L,2019arXiv190103260L}. It is required that the cyclotron frequency be much larger than the wave frequency in the emission region. This requirement limits emission regions to the \mss\  of neutron stars. 

A somewhat separate issue is the escape of the powerful radio waves from the \ms: as the waves propagate in the (presumably)  dipole  \ms\, their amplitude decrease slower than that of the guide field. 
\cite{2021ApJ...922L...7B,2022PhRvL.128y5003B,2023arXiv230712182B} argued that strong \EM\ wave, even if generated with the magnetars' \mss, would not escape.   The argument, in the simplest form, goes as: when the \EM\ field becomes larger than the guide field, for some waves, for which there is a component of the wave's \Bf\ along the guide field, there  are periodic instances when \Ef\ becomes larger than the \Bf. This lead to efficient particle acceleration, and dissipation of the wave's energy.
Along similar line of reasoning, \cite{2023arXiv230909218G}  argued that  the nonlinear decay of the fast magnetosonic into the \Alfven\  waves  would lead to efficient energy dissipation of the wave. 

Here we argue that the particular case considered  in \cite{2021ApJ...922L...7B,2022PhRvL.128y5003B,2023arXiv230712182B} are extreme, not indicative of the more  general situation \citep[see also][]{2022MNRAS.515.2020Q}.  Most importantly, ponderomotive acceleration results in a very slow rate of overtaking the particle by the wave - particles surf the weaker rump-up part of the pulse for a long time, experiencing mild local intensity of the wave, and radiative losses much smaller than in the fully developed pulse.
Same criticism applies to    \cite{2023arXiv230909218G}   -  ponderomotive acceleration would greatly reduce the efficiency of non-linear waves' interaction (by approximately $\sim \gamma_\parallel^3 \gg 1 $). 

Several other   related issues are:  (i) geometry of the \Bf\ and  wave polarization:  \cite{2021ApJ...922L...7B,2022PhRvL.128y5003B,2023arXiv230712182B} considered X-mode (when the \Bf\ of the wave adds/subtracts from the guide field) propagating nearly perpendicularly with respect to the guide field \citep[\eg, strictly equatorial propagation considered in][]{2023arXiv230712182B} 
- this is the most dissipative case.

\section{Non-linear \EM\ waves with guide field}

\subsection{Basic parameters} 
 There  are several important parameters for non-linear wave-particle interaction. First there is the  laser non-linearity parameter  \citep{1975OISNP...1.....A}
\be
a_0 \equiv \frac{e E_w}{m_e c \om}
\label{Akhiezer} 
\ee
where $E_w=B_w$ is the \Ef\ in the coherent wave, and $\om$ is the frequency (parameter $a_0$ is Lorentz invariant).
  In the absence of guide field   the nonlinearity  parameter (\ref{Akhiezer}) is  of the order of a dimensionless momentum of transverse motion of a particle in the EM wave, in the frame where particle is on average at rest. In this case (no guide field), for   $a_0 \geq 1$ the particle motion becomes relativistic. The transverse \Lf, as measured  in the gyration frame is
$
 \gamma_0 = \sqrt{1+a_0^2}
 \label{gammaperp} 
$
 (for circularly polarized waves; $ \gamma_0 = \sqrt{1+a_0^2/2}$ for linearly polarized wave). 
 
 The second important parameter  is the relative  amplitude of the \EM\ field of the wave with respect to the guide field
 \be
 \delta= \frac{B_w}{B_0} 
 \ee
 
 Then, there is the ratio of the cyclotron frequency to wave frequency 
  \be
 f=  \frac{\om_B}{\om}
 \ee
 The three parameters combine 
 \be
 a_0  =  \delta f
 \ee
 The corresponding combination on the rhs is  Lorentz invariant under a boost along  the guide \Bf.
 
It turns out, see Eq. (\ref{p0ofa0}), that  another  important combination is 
\be
\tilde{a}_0 = \frac{a_0}{ 1+f}
\label{tildea0}
\ee
This is effective non-linearity parameter for non-linear \EM\ wave in finite guide field (parameter $f$ is defined in the lab frame, where initially a particle is at rest.).

   

\subsection{FRBs' parameters}

For fiducial estimates,  consider an FRB  pulse coming from $d$= Gpc, of duration $\tau = 1$ msec,  and producing flux $F_\nu$ = 1 Jy  at frequency of  $\nu=10^9$ Hz (these values are at the higher end of the FRB parameters). The isotropic equivalent luminosity  and total energy (in radio)  are then
\ba &&
L_{iso} =  4 \pi D^2 \nu F_\nu  = 10^{42} {\rm erg\, s}^{-1}
\nn &&
E_{iso} = L_{iso} \tau =  10^{39} {\rm erg}
\label{Eiso}
\ea
The \EM\ field of the wave at distance $r$ from the source is 
\be
B_w =  \frac{  2  \sqrt{\pi}  \sqrt{\nu F_\nu}    d}{ \sqrt{c} r  } = 6 \times 10^{9}   \left( \frac{ r}{ R_{NS}}  \right) ^{-1}  \, {\rm G}
\label{Eiso1}
\ee

The laser non-linearity parameter then evaluates to 
\be
 a_{0}^\ast  = \frac{ e \sqrt{F_\nu}  d}{\sqrt{\pi}  m_e c^{3/2} \nu^{1/2} R_{NS} } =  10^7
\ee

\begin{figure}[h!]
\includegraphics[width=0.99\linewidth]{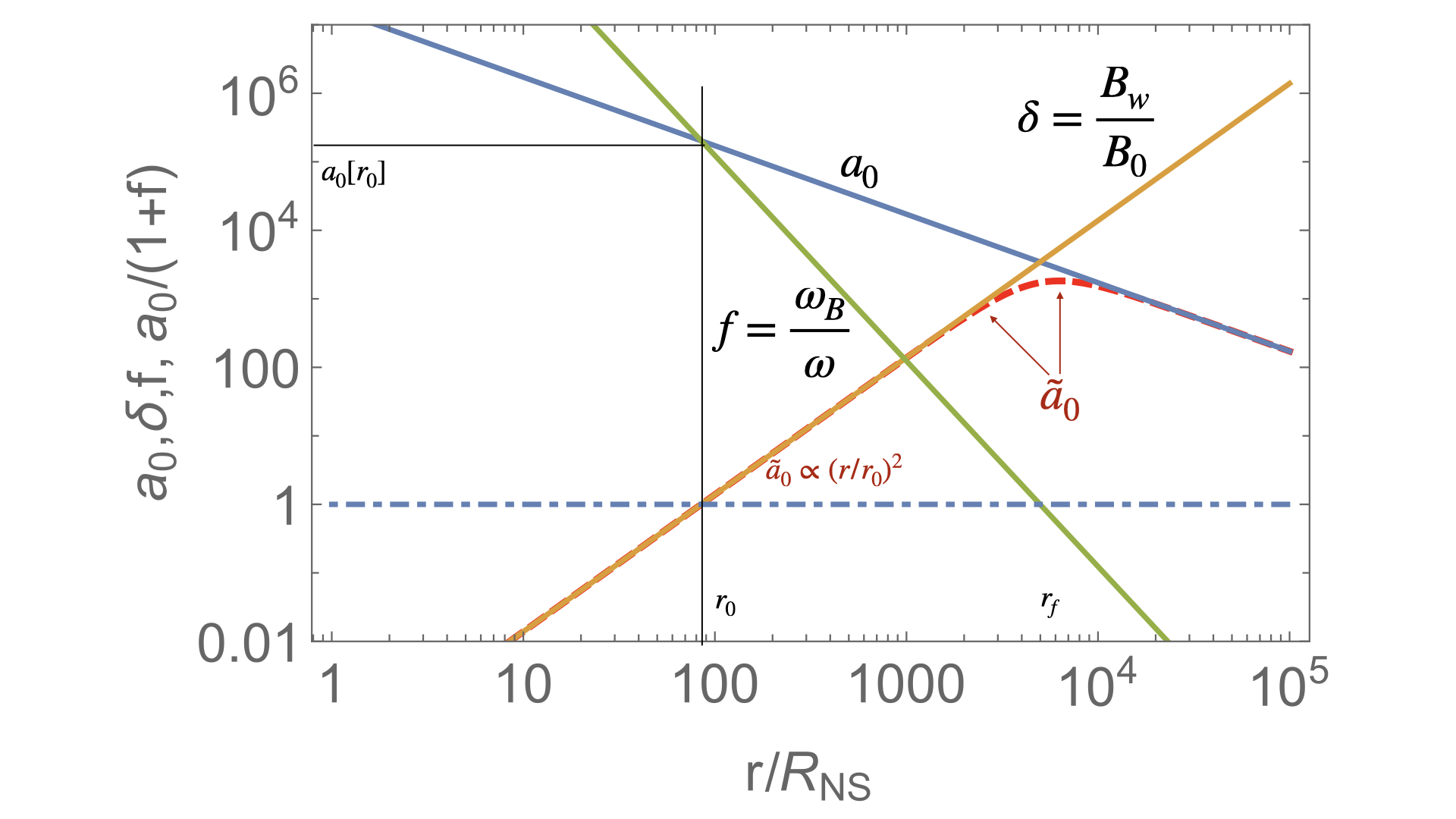}
\caption{
 Evolution of basic parameters in the dipolar \ms: nonlinearity parameter $a_0$, relative wave intensity $\delta = B_w/B_0$, ratio of frequencies $f= \om_B/\om$,  and effective   nonlinearity parameter $\tilde{a}_0 = a_0/(1+f)$. Indicated are radii $r_0$ (where  $\tilde{a}_0=1$), value of $a_0[r_0]$, and radius $r_f$ where $f=1$.  Maximal value of  $\tilde{a}_0 ^{(max)} \approx 1.8 \times 10^3 $ is reached approximately at $r_f$.}
\label{FRB-escapeLC}
\end {figure}

 
If we normalize the  surface \Bf\ to the quantum field
\ba && 
B_{NS} = b_q  B_Q 
\nn &&
B_Q =  \frac{c^3 m_e^2}{e \hbar }
\ea
and for now assume dipolar field
\ba && 
B_0 = B_{NS} (r/R_{NS} )^{-3}
\nn &&
a_0 =  a_{0}^\ast   (r/R_{NS} )^{-1},
\label{BQ}
\ea
then the relative amplitude of the fluctuating field and the ratio of frequencies are
\ba && 
\delta = \frac{ B_w}{B_0}=  \frac{ 2  \sqrt{\pi}  d  \sqrt{\nu F_\nu}  r^2 }{\sqrt{c}  B_{NS} R_{NS} } =  10^{-4}  b_q ^{-1}  \left( \frac{ r}{ R_{NS}}  \right) ^{2} 
\nn &&
f= b_q \frac{ e B_Q}{2\pi m_e c \nu}   \left( \frac{ r}{ R_{NS}}  \right) ^{-3}=10^{11} b_q  \left( \frac{ r}{ R_{NS}}  \right) ^{-3} 
\ea
 see Figs. \ref{FRB-escapeLC}.
 
The amplitude of \EM\ fluctuations in the wave becomes comparable to the guide field ($ \delta \sim1 $) at
\be
\frac{r_0}{R_{NS}} \approx   10^2  b_q^{1/2}
\label{r0}
\ee
At that point
\ba && 
a_0 [r_0] \approx 2\times 10^5 
\nn &&
f (r_0) = a_0 [r_0]= 2 \times 10^{5} 
\label{aoofr0}
\ea
The wave's frequency equals cyclotron frequency at
\ba &&
\frac{r_f}{r_0} = a_0 [r_0] ^{1/3}
\nn &&
\frac{r_f}{R_{NS}} \approx 5 \times 10^3  b_q^{1/3} 
\nn &&
a_0(r_f) = a_0 [r_0]^{2/3}
\label{rf} 
\ea

The key parameter $\tilde{a}_0$, Eq.  (\ref{tildea0}),  has a maximum  (for dipole field) at 
\ba &&
r/r_0= (2 a_0 [r_0])^{1/3}= 2^{1/3} r_f
\nn &&
 \tilde{a}_0^{(max)} =  \frac{2^{2/3}}{ 3} a_0 [r_0]^{2/3}
\ea
This estimates to
\ba &&
 \left( \frac{ r}{ R_{NS}}  \right) = b_q^{1/3}  \left(\frac{c^2 m_e}{\pi   \hbar  \nu }\right) ^{1/3} = 6 \times 10^3 \times   b_q^{1/3}
 \nn &&
 \tilde{a}_0^{(max)} \approx 1.8 \times 10^3\times  b_q^{-1/3}
 \label{aomax}
 \ea

Parameter $  \tilde{a}_0^{(max)} $, Eq. (\ref{aomax}) is an important one.  This is an estimate of the maximal total  \Lf, $\sim  \tilde{a}_0^{(max), 2}$, maximal parallel   \Lf, $\sim  \tilde{a}_0^{(max)}$,  and maximal transverse momentum   $\sim  \tilde{a}_0^{(max)}$.

For example,  maximal values of  the \Lf\ are achieved at $ r/r_0 = (2 a_0 [r_0])^{1/3}  $ and equals 
\be
\gamma_{p} = 1 + \frac{2^{1/3}  }{9}a_0 [r_0]^{4/3}
\label{gammap}
\ee
These values are large, and in the \mss\ of magnetars would lead to large radiative losses, killing the EM pulse. We argue that these high values are not reached.

\begin{figure}[h!]
\includegraphics[width=0.99\linewidth]{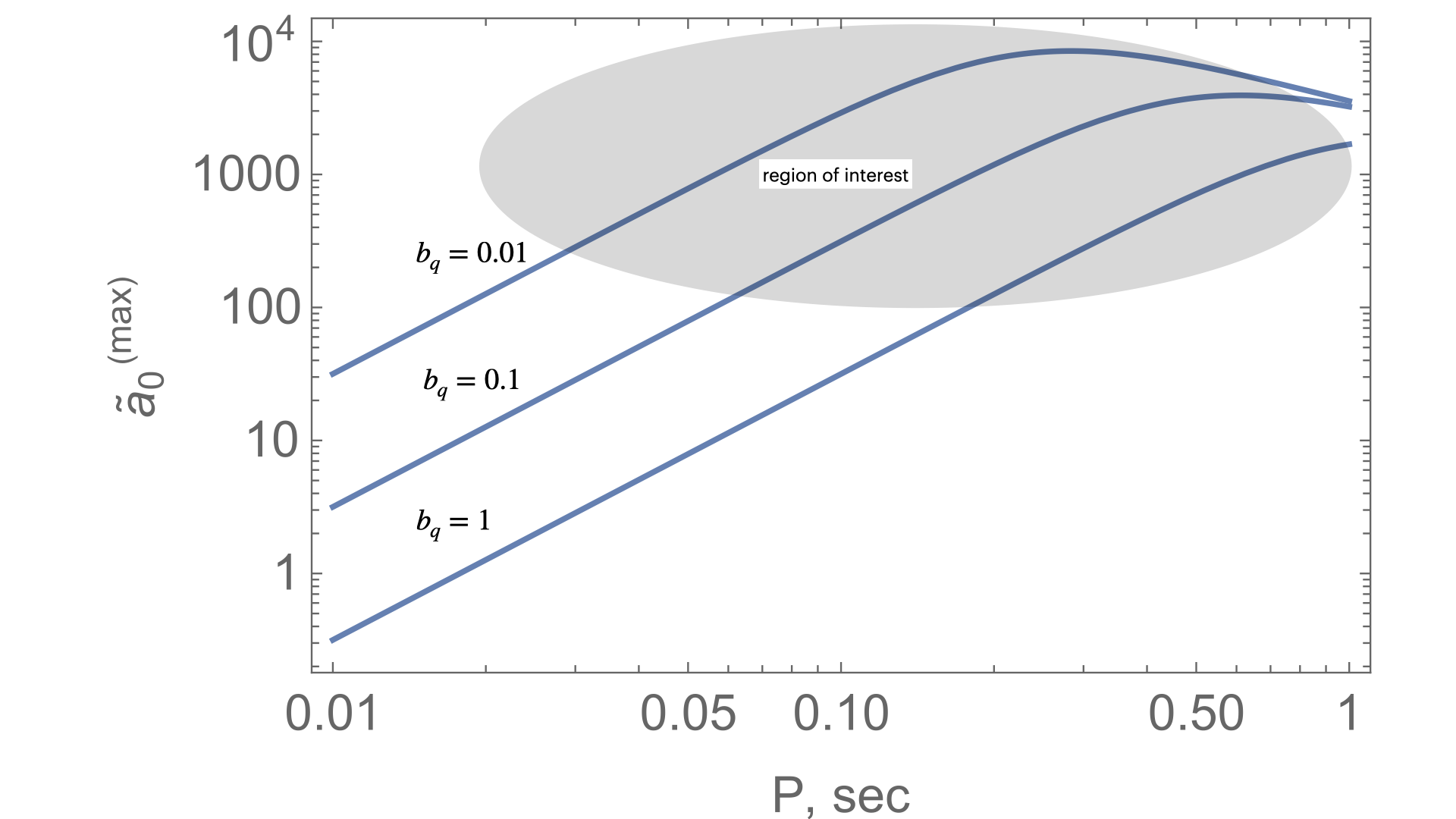}
\caption{
 Maximal values of the non-linearity parameter  $ \tilde{a}_0$ reached at the \LC\ for spins faster than $P=1$ second (for slower spins see Eq. (\ref{aomax})), as function of the surface field $b_q=B_{NS}/B_Q$. Smaller magnetic  fields are less  efficient in suppressing $ \tilde{a}_0$. "Region of interest" indicates a set of parameters when wave's nonlinearity may become large, $\geq 100$.}
\label{aomaxofP1}
\end {figure}

For the maximal $ \tilde{a}_0^{(max)} $ to be reached with the \ms, the period should be sufficiently long $P \geq 1$ sec.   For shorter periods the value of  $ \tilde{a}_0^{(max)} $ is reached at the \LC, see (\ref{aomaxofP1}).  For mildly magnetized \NSs, with regular surface field  $\sim 10^{12}$ G ($b_q=0.02$), spinning with period $P \sim 20 $  milliseconds, the value of $ \tilde{a}_0^{(max)} $ can reach nearly $10^4$.  It is not clear why FRB sources would fall into this special regime: higher surface field and faster spins push $ \tilde{a}_0^{(max)} $ towards smaller values.  For example, for quantum surface field $b_q=1$ and period of 10 milliseconds,   $ \tilde{a}_0^{(max)} $ is tiny, $\leq 1$. 

\subsection{Post-eruption magnetic field lines are mostly radial, \Bf\ increases in the outer parts of the \ms}
\label{Post-eruption}

There are, qualitatively, three energy sources during FRB/magnetospheric eruption: (i) dynamical \Bf; (ii)  high energy  emission; (iii)  radio emission. Energetically, (i) $\gg$ (ii)  $\gg$ (iii), so that the dominant effects to the distortion of the \Bf\ come not from radiation, but from magnetospheric dynamics during the eruption.

The dynamical \Bf\ is produced by a process that initiates magnetospheric eruption, \eg, in an analogue to Solar Coronal Mass Ejection (CME). During CME a topologically isolated structure is injected  into the \ms, \cite[][]{2022MNRAS.509.2689L,2023MNRAS.524.6024S}.
Let's assume that injection (generation of topologically disconnected magnetic structure)  occurs near the stellar surface with 
the   typical size $R_{CME,0} \leq R_{NS}$ and associated energy $E_{CME,0}$.
An important parameter is the total magnetic energy of the \ms,
\be
E_{B, NS} \sim B_0^2 R_{NS}^3
\ee
Naturally, the injected energy is much smaller than the total energy,
\be
\eta_{CME} = \frac{ E_{CME,0}}{E_{B, NS}} \leq 1
\ee

As the CME is  breaking-out through the overlaying \Bf, it does work on the magnetospheric \Bf. At some point ``detonation'' occurs: when the total energy contained in the confining \Bf\ exterior to the position of the CME ($\sim B_0^2 R_{NS} ^6 r^{-3}$)  becomes smaller than the CME's internal energy (equivalently, when the size of the CME becomes comparable to the distance to the star).  This occurs at some equipartition radius $r_{eq}$:
\be
\frac{r_{eq} } {R_{NS}}  \sim
\frac{E_{B, NS}} {E_{CME,0}} = \eta_{CME}^{-1} \geq 1
\label{req} 
\ee
Immediately after the generation of a CME the \ms\ becomes open, with nearly radial \Bf\ lines for $r\geq r_{eq} $. 

For quantum surface field $E_{B, NS} \sim 2 \times 10^{45}$ erg. The injected energy $E_{CME,0}$ is hard to estimate: the observed radio energy is an absolute lower limit. Much more energy is radiated in X-rays, even more is contained in the fields. \cite[Also CME is losing energy to $pdV$ work as it breaks through the overlaying \Bf][]{2022MNRAS.509.2689L,2023MNRAS.524.6024S}.
It is conceivable that the relative injected energy may reach $\sim$ percent level of the total energy, $\eta_{CME} \sim 10^{-2}$. In this case, since beyond  $r_{eq}$ the \Bf\ decreases slower than dipole, $B\propto r^{-2}$ instead of  $B\propto r^{-3}$, the  the region where $f   =\om_B/\om \geq 1$ will extend further. (Larger guide field suppresses particle's transverse motion, see Fig.  \ref{Crab-Alfven-turb-pofa0}.)

 \begin{figure}[h!]
  \centering
    \includegraphics[width=0.8\textwidth]{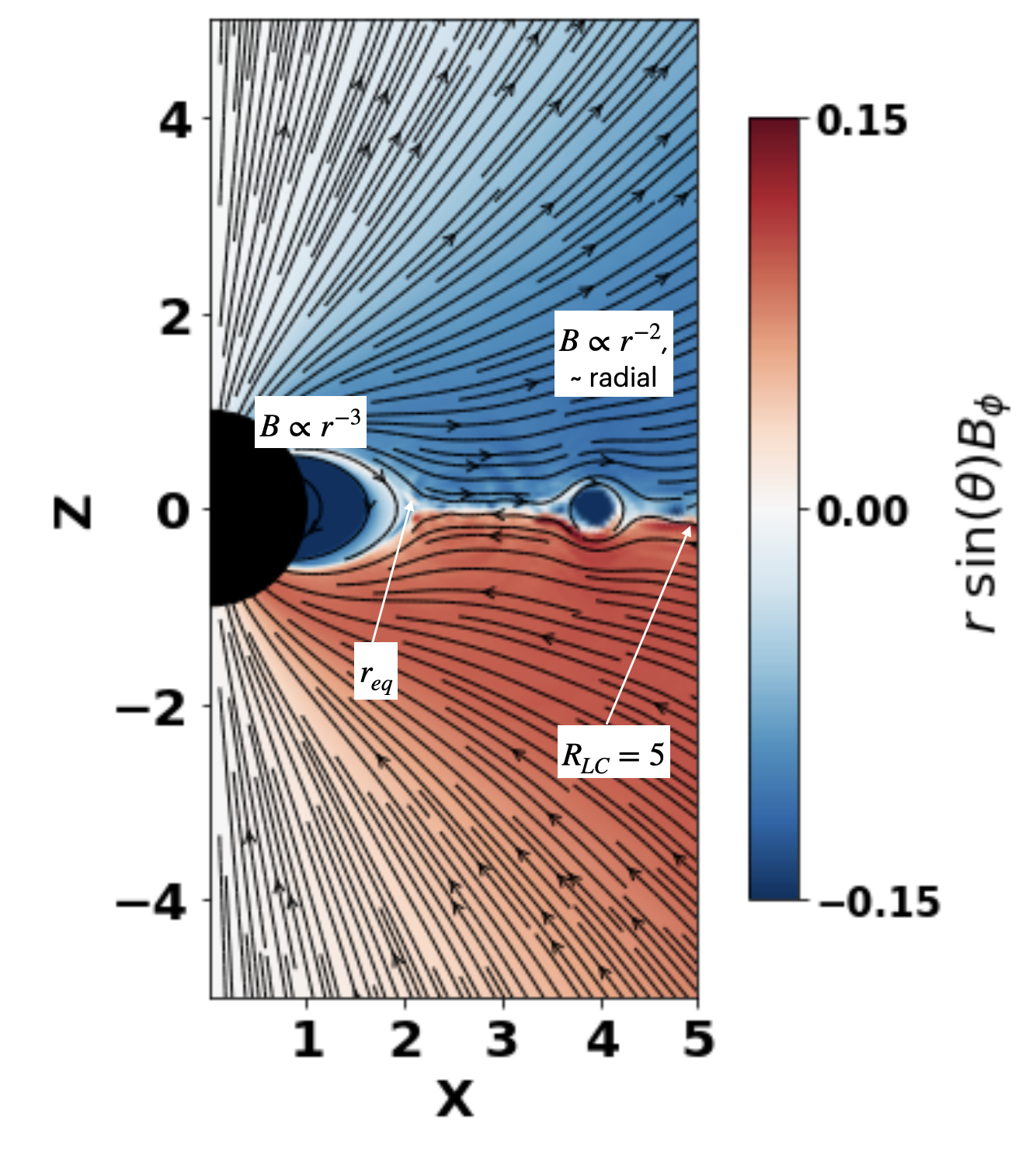}
     \caption{Post-flare opening of the \ms\ \citep{2023MNRAS.524.6024S}. Color is values  $r \sin \theta B_{\phi}$, lines are poloidal field line. The spin parameter is $\Omega =0.2$, so that the \LC\ is at $R_{LC} =5$. Post-flare \ms\ is open starting $r_{eq}  \ll  R_{LC}$, (\ref{req}). Beyond  $r_{eq} $ the \ms\ has  monopolar-like \Bf\ structure. }
\label{StreamlinexBphi_LocalShearing_Symmetric_Pi4}
\end{figure}

Most importantly,   beyond $r_{eq}$ the \Bf\ becomes mostly radial. As a result, the \EM\ waves generated close to the \NS\ surface propagate nearly along the local field line. As we demonstrate in this work, this case does not suffer from  strong radiative.

Finally,  plasma is likely to stream out along the open field lines even before the FRB wave comes -  this further freezes out wave-particle interaction, see Fig. \ref{gammaoft}.

\section{Particle dynamics in circularly polarized wave propagating along  guide field}

\subsection{Beam frame}
\label{special}

Circularly polarized waves allow for exact analytical solutions, and thus provide benchmark for simulations and guidance for the more complicated linearly polarized  case.

In the beam frame a force balance for a particle moving in \EM\ field and guide \Bf\  reads 
\be
\gamma_\pm m_e v _\pm \om= e ( E_w \pm  v _\pm B_0)
\label{Forces}
\ee 
where   all quantities are positive: $ \pm $ accounts for two directions of the background field/charge/polarization sign (speed of light is set to unity). 
Relations describe a charge which velocity  at each moment (counter)-aligns with the \Bf\ in the wave  \citep{1975UsFiN.115..161Z}. For a more general case see \cite{1964PhRv..135..381R,2007PhPl...14f3101K}.

\begin{figure}[h!]
\includegraphics[width=.99\linewidth]{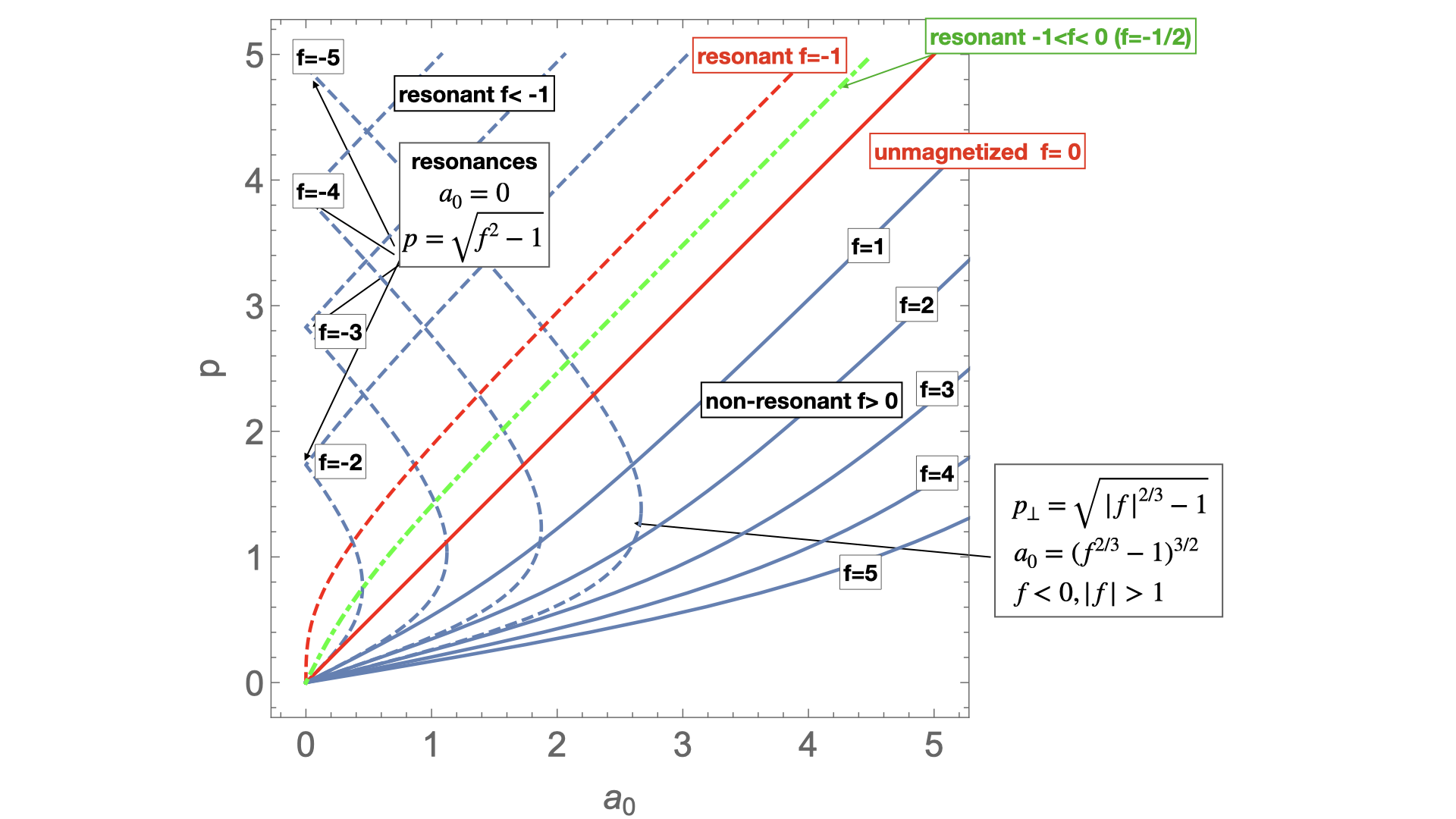}
\caption{Particle's transverse  momenta in strong circularly polarized  \EM\ wave in external \Bf. Wave intensity is parameterized by laser parameter $a_0$. Different curves correspond to different $f= \pm  \om_B/\om$ (different charges or polarizations). } 
\label{Crab-Alfven-turb-pofa0}
\end{figure}

In dimensionless notations  the motion of a particle in circularly polarized  \EM\ wave obeys
\ba&&
a_0=  p_0   \left|1+ \frac{1}{\sqrt{1+p_0^2}}  f \right| 
\nn &&
\gamma_0 = \frac{1}{\sqrt{1-v_0^2}}
\label{a00}
\ea
(here  $\om$ is the wave  frequency in the gyration frame), see Fig. \ref{Crab-Alfven-turb-pofa0}. 
Quantity $f$ can be negative: two signs correspond to two polarizations (or two signs of charges). Absolute value $|...|$ ensures  the definition of $a_0 \ge 0$; crossing the resonant condition for the minus sign  changes just the phases of the particles. Below in this section  we drop the prime, with clear understanding that the quantities are measured in the plasma frame.

In (\ref{a00}) the values for $\gamma_0$ (for given $a_0$ and $|f|$) are different for two charges, especially near the cyclotron resonance $f \approx -1$. As we  are not considering here  the effects of cyclotron absorption,  we assume below that $- f$ is not too close to unity.


In the case  of strong \EM\ pulse with $a_0 \gg  1$, there are qualitatively three  regimes: (i) small guide field $f \ll1$, $p_0 \sim a_0$; (ii) medium  guide field $ 1 \ll f \ll a_0 $, $p_0 \approx \delta \gg1$ (in this case the transverse motion is still relativistic: a wave is sufficiently strong so that it accelerate particles to relativistic motion on time scale of $1/\om_B$ ;  (iii) dominant guide field $f \gg a_0$, $v_0 \sim \delta \ll 1$ (in this case a particle just experiences E-cross-B drift). 
 
  \subsection{Ponderomotive acceleration by circularly polarized wave propagating along the guide field}
   \label{pondero}
  The above discussion in \S \ref{special} omits the most important issue: the ponderomotive effects - how incoming wave modifies the properties of the plasma.  As this is the most important part of the work, we give here detailed derivations.


  Let a transverse  circularly polarized wave of given strength $E_w$, frequency $\om$ (measured in lab frame), non-linearity parameter $a_0$,  propagating along guide \Bf\  $B_0 {\bf e}_z$.
  Noting that 
  \ba &&
  \partial_ t \gamma = \frac{e }{m_ e c^2} {\bf E}\cdot{\bf v}
  \nn &&
   \partial_ t p_z = \frac{e} {c}  \left. {\bf v} \times {\bf B}_w\right|_z
 \ea
 and expressing  fields in terms of the vector potential
 \ba &&
 \E_w = - \partial _t {\bf A}
 \nn &&
 \B_w= \curl {\bf A}
 \ea
 we find
 \ba &&
 \partial_t \gamma = -v_x \partial_t A_x - v_y  \partial_t A_y
 \nn &&
  \partial_t p_z  = v_x \partial_z A_x + v_y  \partial_z A_y
  \ea
  Thus, the guide field does not enter the relations. Since $A=A(z-t)$,  we find then
 \be
 d_t( \gamma - p_z m_e c) =0
 \label{para1}
 \ee
 Switching to dimensionless notations and assuming that before the arrival of the wave a particle was at rest, we find
 \be
 \gamma =1+p_z
  \label{para11}
 \ee
 We stress that for circularly polarized wave propagating along the \Bf,  this is valid for arbitrary guide field.
 
 Thus (recall that $p_0$ is the transverse momentum, hence Lorentz invariant under a boost along $z$)
 \ba && 
   p_z = \frac{{\bf p}_0^2}{2} 
   \nn && 
   \gamma =1+p_0^2/2 =1+p_z
   \nn  &&
    \gamma_\parallel =\frac{1}{\sqrt{1-\beta_z^2}} = \frac{1+p_0^2/2}{\sqrt{1+p_0^2}} = \frac{1+p_z}{\sqrt{1+2 p_z}}
    \nn &&
    \gamma_0 = \sqrt{1+p_0^2}
 \nn && 
   \beta_z = \frac{p_0^2}{2+ p_0^2}
     \nn &&
     \tan \alpha_p = \frac{ p_0}{ p_z} = \frac{2} { p_0}
     \label{44}
\ea
where $\alpha_p$ is pitch angle in lab frame. (Note that $\gamma_\parallel \neq \sqrt{1+p_z^2}$.)
 These relations establish connection between parallel motion acquired due to ponderomotive force and energy of the particle in the gyration/beam frame for circularly polarized wave, possibly  propagating along   guide \Bf.
 
 One remaining step is to connect $p_z$ (or $p_0$) to the waves' parameters  $a_0$ and $f$ at minus infinity, before interaction with a particle.
 Using invariance of $a_0 = E_w/\om$ and $\om_B$, and  Lorentz transformation of the frequency 
\be
 \om ' =(1- \beta_\parallel) \gamma_\parallel \om
 \label{omres}
\ee
(where now primes denote quantities measured in the beam frame)
we    arrive at
\ba &&
p_0 = \frac{a_0}{1+f}
\nn &&
\gamma_\perp = \sqrt{1+p_0^2}=  \sqrt{1+\left(  \frac{a_0}{1+f} \right)^2} 
\nn && 
p_z =  \frac{a_0^2}{2(1+f)^2}
\nn &&
\gamma = 1+ \frac{a_0^2}{2(1+f)^2}
\nn &&
\beta_z = \frac{a_0^2}{2 (1+f) ^2 +a_0^2}
\nn &&
  \gamma_\parallel = \frac{1}{\sqrt{1-\beta_z^2}}=   \frac{a_0^2+2 (1+f)^2}{2 (1+f) \sqrt{a_0^2+(1+f)^2}} = \frac{\gamma}{\gamma_\perp}
\nn  &&
\tan  \alpha = \frac{p_0}{p_z} = \frac{2 (1+f)}{a_0}
 \label{p0ofa0} 
 \ea
 
 We observe that the case with guide field is related to the no-guide field if we use 
\be
\tilde{a}_0= \frac{a_0} {1+f} 
\ee
Then relations in \Bf\ reduce to the same form as without guide field
\ba &&
p_0 = \tilde{a}_0 
\nn && 
p_z =  \frac{ \tilde{a}_0^2}{2}
\nn &&
\gamma= 1 +   \frac{ \tilde{a}_0^2}{2}
\nn &&
\tan  \alpha = \frac{2 }{\tilde{a}_0}
 \label{p0ofa01} 
 \ea

Importantly, relations (\ref{p0ofa0}-\ref{p0ofa01}) assume that the system  is sufficiently large along the direction of propagation, so that a particle reaches the final steady state. As we show below, this is often not the case in magnetars' \mss.

 \begin{figure}[h!]
\includegraphics[width=.49\linewidth]{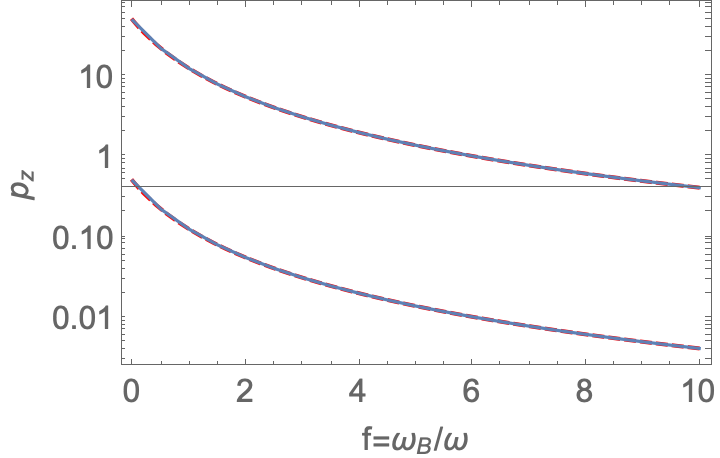}
\includegraphics[width=.49\linewidth]{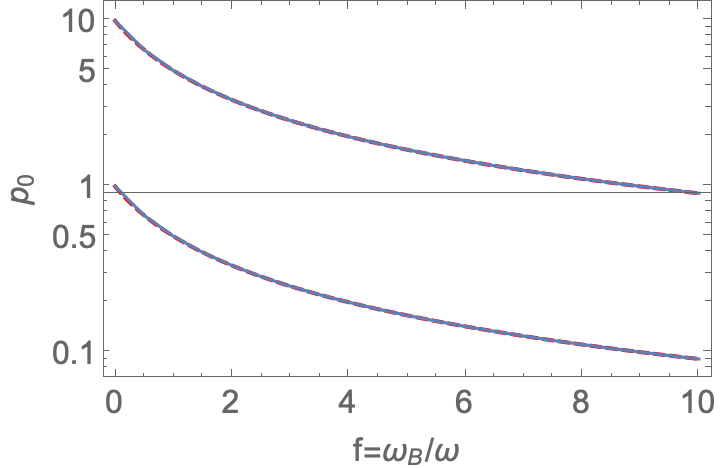}
\caption{Ponderomotive acceleration by a circularly polarized \EM\  pulse propagating along the guide field, non resonant case $f>0$. Plotted is the axial momentum $p_z$  (left panel) and transverse momentum $p_0$  (right panel) as a function of $f= \om_B/\om$ ($\om$ defined in the  lab frame) for two cases: $a_0 =1$ (bottom curves) and $a_0=10$  (top curves).  Dashed line is analytical result, Eq. (\ref{p0ofa0}). Numerical and analytical curves nearly coincide. These plots serves to illustrate the numerical precision of the code, and to validate the analytics.} 
\label{pzoff}
\end{figure}

   \subsection{Ponderomotive surfing in constant guide field}
 
 Ponderomotive force has another important effect: in a system limited in size, the head part of the pulse, which is already non-linear but has local nonlinearity parameter much smaller than the pulse,   will accelerate a particle to relativistic velocities along wave's direction of propagation, so that it will take a long time for the bulk of the pulse to catch-up with the particle. We consider this effect next.
   
\begin{figure}[h!]
\includegraphics[width=.49\linewidth]{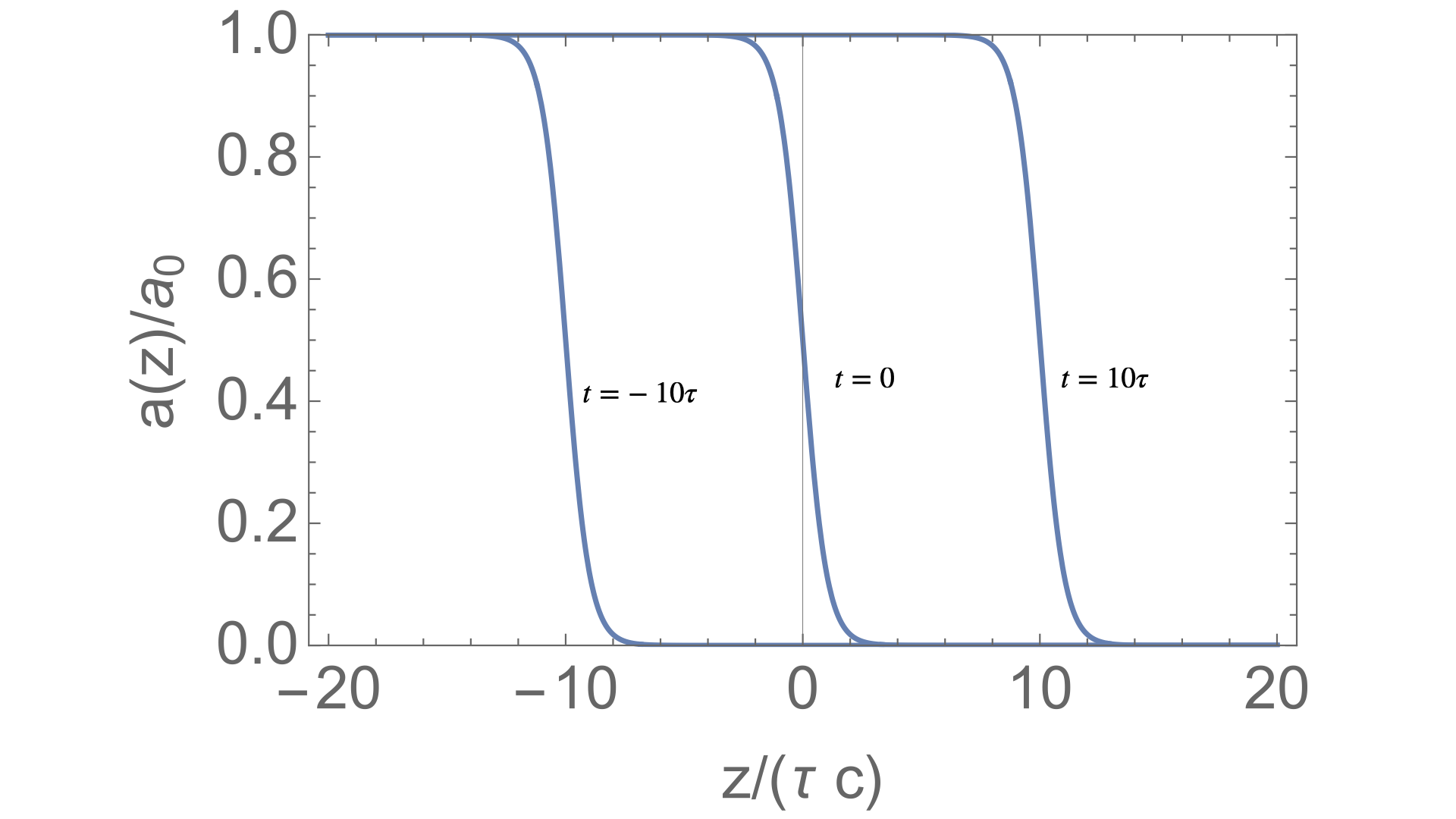}
\includegraphics[width=.49\linewidth]{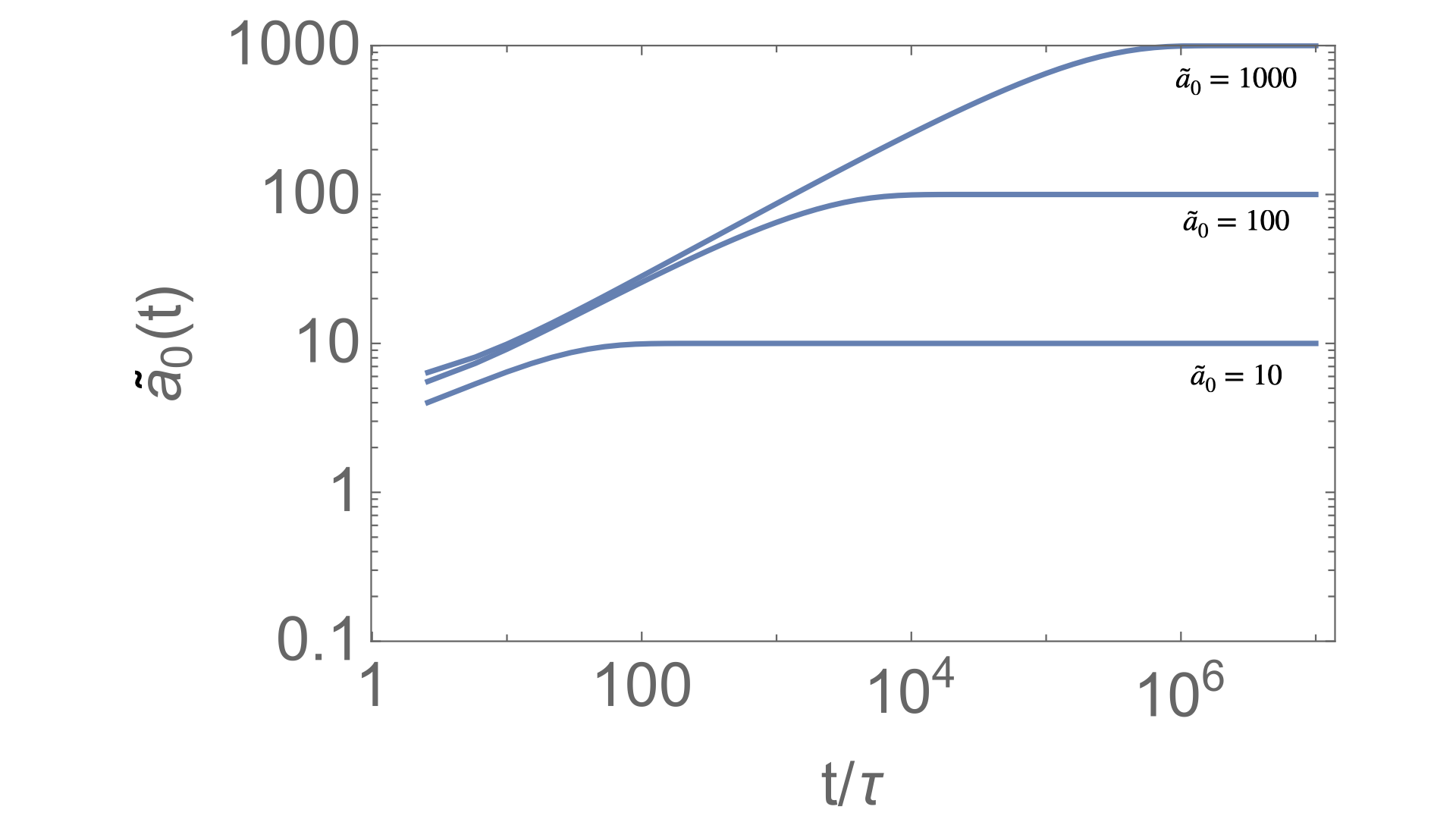}
\caption{Left panel: An EM  pulse  with ramp-up width of $\tau =1$ propagates towards a particle initially at $z=0$. The pulse ponderomotively accelerates the particle along its direction of propagation.  Right panel: Evolution of the non-linearity parameter $\tilde{a}_0(t)$ {\it  at the location of the particles} for $\tilde{a}_0 =10,\, 100,\, 1000$, constant guide field. Due to ponderomotive acceleration of the particle  to $\gamma_\parallel \sim \tilde{a}_0$, the bulk of the pulse reaches the particle after a very long time,  $\sim  \tilde{a}_0^2  \tau$.}
\label{aofz}
\end{figure}

   Consider a wave propagating along the field (see \S \ref{Self-cleaning} for oblique case.) 
  Assume that a pulse approaches a particle initially at rest at $z=0$. The pulse has maximal amplitude $a_0$ and ramp-up width $\delta z = \tau c$, Fig. \ref{aofz}. At each moment the axial velocity is 
   \be
   \beta_z = \frac{d z}{dt}=  \frac{\tilde{a}(z)^2}{2+\tilde{a}(z)^2}
   \label{betaz} 
   \ee
   where $\tilde{a}(z)$ is the wave amplitude at the current location of the particle, Fig. \ref{aofz}. Equation (\ref{betaz}) can be integrated for $z(t)$ assuming some given profile of the pulse (\eg $\tanh(z-t) / \tau$).

   What is important is not only the absolute value of the intensity $\tilde{a}_0$, but also temporal evolution of the non-linearity parameter {\it at the location of the particle}, Fig. \ref{aofz} right panel. Due to ponderomotive acceleration of the particle  to $\gamma_\parallel \sim \tilde{a}_0$, the bulk of the pulse reaches the particle after a very long time,  $\sim (2/5) \tilde{a}_0^2  \tau$. 
   
\begin{figure}[h!]
\includegraphics[width=.49\linewidth]{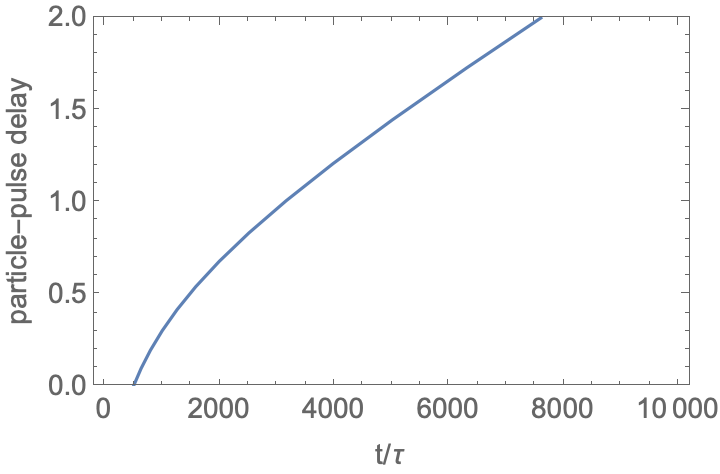}
\includegraphics[width=.49\linewidth]{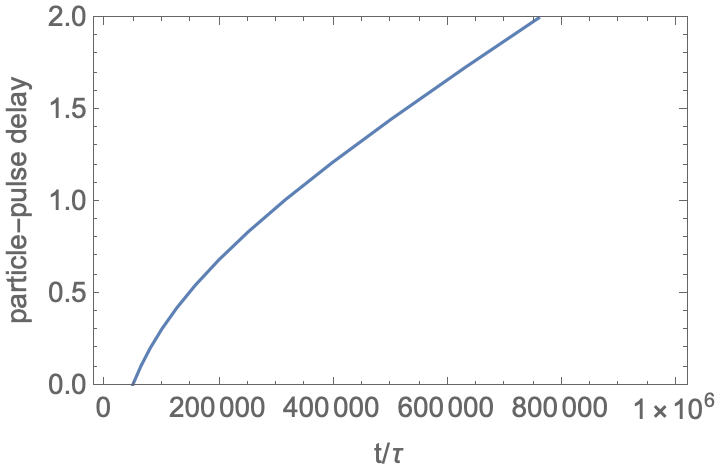}
\caption{Pulse front - particle delay as function of time (measured in terms of pulse ramp-up time $ \tau$ for $\tilde{a}_0=100$ (left) and   $\tilde{a}_0=1000$ (right), constant guide field. The delay becomes of the order of unity after time $\sim (2/5) \tilde{a}_0^2  \tau$. }
\label{delay}
\end{figure}

   As another measure, in Fig. \ref{delay} we plot a delay between the particle and the center of the pulse (located at $z= c t$).  The delay becomes of the order of the width after time  $\sim (2/5) \tilde{a}_0^2  \tau$. (The center of the pulse, where local nonlinearity parameter is $\tilde{a}_0/2$ does not even overtake a particle before time $\approx  \tau \gamma_0^2/10$.)

\subsection{Adiabatic force} 
The above relations omit an important effect: the adiabatic force that accelerates particles along decreasing \Bf\ at the expense of transverse motion.

   Adiabatic force  can be written as (the parallel component is sometimes called the  mirror force)
   \ba &&
G_{\phi} = \pm \frac{\beta^2 \gamma m_e c^2}{4}   \frac{  ({\bf b} \nabla) B}{B} \sin 2 \alpha 
\nn &&
G_{z} = \mp \frac{\beta^2 \gamma m_e c^2}{2}   \frac{  ({\bf b} \nabla) B}{B} \sin ^2 \alpha 
\label{G} 
\ea
where ${\bf b} $ is unit vector along the \Bf, $\alpha$ is pitch angle,  and upper (lower) signs correspond to particle propagating towards (away from) the regions of increasing \Bf. 
Since ${\bf G} \cdot {\bf v}=0$, the \Lf\ $\gamma=$ constant.  Adiabatic force can be thought of as $({\bf m} \cdot \nabla)  \B$ force, where   ${\bf m} $ is the  magnetic  momentum. 

Neglecting  particular dependence of the value of the \Bf\ on  the polar angle,
\be
\frac{({\bf b} \nabla) B}{B} = -\frac{1}{3r}
\ee
Where  we chose axis $z$  along the local \Bf,   pointed away from the star, and we assumed that particles are   moving away.

In dimensionless notations
\ba && 
\partial_t p_\perp = -\frac{\beta ^2 \gamma  p_\perp p_z}{6 \left(\gamma ^2-1\right) r}
\nn &&
\partial_t p_z= \frac{\beta ^2 \gamma  p_z^2}{6 \left(\gamma ^2-1\right) r}
\ea
Adiabatic force accelerates along the field at the expense of transverse motion.

\begin{figure}[h!]
\includegraphics[width=.99\linewidth]{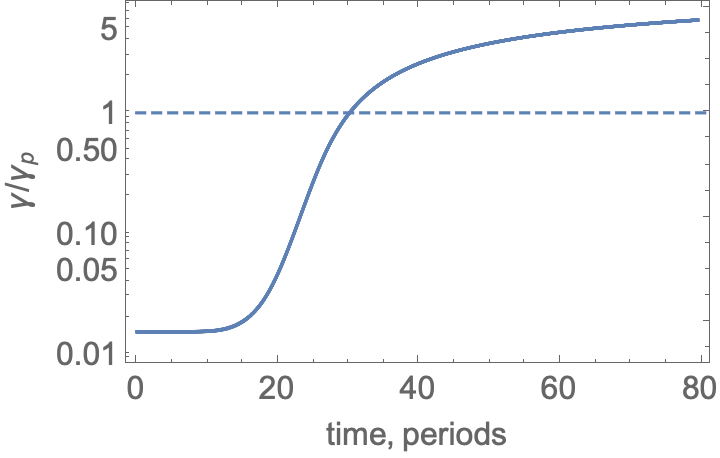}
\caption{Effect of adiabatic acceleration by Gaussian pulse. Dashed curve: expected maximal \Lf\ without effects of adiabatic acceleration, Eq. (\ref{gammap}). This illustrates the importance of  adiabatic acceleration. Initially, $a_0 = f=100$, so $\gamma_p =66$. At time $t=0$ the peak of the pulse is at $-20$ wavelengths, rise time is $5$ wavelengths, scale of \Bf\ decrease is 10 times the   rise time. This confirms that adiabatic effects have about $10\% $ influence on particle dynamics} 
\label{adiabat-acc}
\end{figure}

To get a feeling of how  the adiabatic force affects the dynamics, let' us assume that it acts on time scales longer than pulse ramp-up time. In this case, at each radius there is ponderomotive force, so that total force balance reads 
\ba && 
\partial_r p_\perp = -\frac{\beta ^2 \gamma  p_\perp p_z}{6 \left(\gamma ^2-1\right) r} + \partial_r \tilde{a}_0
\nn &&
\partial_r p_z= \frac{\beta ^2 \gamma  p_z^2}{6 \left(\gamma ^2-1\right) r} + \tilde{a}_0 \partial_r \tilde{a}_0
\ea
In the region $ r_0\ll r\ll  r_f$, $\tilde{a}_0 \approx (r/r_0)^2$. Assuming relativistic motion $\gamma \gg 1$, and $p_z \gg p_\perp$
\ba && 
\partial_r p_\perp =  - \frac{p_\perp}{6 r} + \frac{2 r}{r_0^2} 
\nn &&
\partial_r p_z=\frac{p_z}{6 r}   + \frac{2 r^3}{r_0^4} 
\ea
with solutions
\ba && 
 p_\perp = \frac{12}{13} \frac{ r^2}{r_0^2} 
 \nn && 
 p_z=   \frac{12}{23} \frac{ r^4}{r_0^4} 
\ea
Thus,  in the regime of short pulses, with ramp-up scale $\ll r$, the adiabatic force has $\sim 10\%$ effect on the particle dynamics, increasing parallel momentum and decreasing the transverse one. Qualitatively, since the losses are high powers of the transverse momentum, this will reduce the losses by about 50\%. Our numerical results confirm this conclusion, Fig. \ref{adiabat-acc}.

 The adiabatic force  helps somewhat particles to avoid losses, as it  decrease transverse momentum (hence decreasing radiative losses) and increases parallel momentum (hence increasing the surfing time). 
   
   \section{``Gone  with the pulse''}

Above we separately described various ingredients - overall particle dynamics in the beam frame,  ponderomotive  and adiabatic accelerations. Next we use these  results  to study particle dynamics within  magnetars' \mss.

  \subsection{Ponderomotive acceleration in magnetars' \mss}
\label{Ponderom} 
First, we take semi-analytical account of ponderomotive acceleratrion  in magnetars' \mss. We solve Eq. (\ref{betaz}) taking into account both the structure of the pulse and spacial dependence of the parameter $\tilde {a_0}$. The following procedure is applied
\begin{itemize}
\item particle is seeded at a given radius
\item An \EM\ pulse of circularly polarized wave is launched from a much smaller radius. The pulse has a  $\propto  \tanh (r-t)/\tau$ profile  with rump-up time.
\item Pulse normalization follows evolution of the parameter $\tilde{a}_0$. 
\item  We numerically integrate Eq. (\ref{betaz}) for the location $z(t)$
\end{itemize}

\begin{figure}[h!]
\includegraphics[width=.99\linewidth]{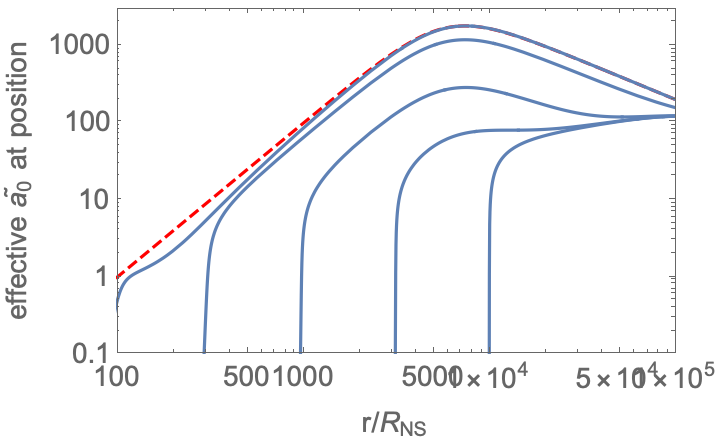}
\caption{Non-linearity parameter $\tilde{a}_0$  at the location of a particle in dipolar-like \ms\  for different initial positions of a particle, subject to a  pulse $\propto  \tanh (r-t)/\tau$ with rump-up time  of  $\tau=10 R_{NS}/c$. Red  line: local value of $\tilde{a}_0$. This illustrates that particles initially located at $\geq$ few $r_0$ do not experience full waves' nonlinearity do to effect of surfing.}
\label{aoefftanh}
\end{figure}

 In 
   Fig. \ref{aoefftanh} we show  evolution of the  $\tilde {a_0}$ in the particle frame (as measured at the location of the particle) for different initial position of the particles. Particles  located close to $r_0 =100 R_{NS}$  initially  experience mild  parallel acceleration, hence quickly overtaken by the head of the pulse and  find themselves in the strong region of the wave (they do not experience much surfing on the rising part). Particles starting  further out quickly gain large {\Lf}s, surf the front part of the pulse, and never experiences near-maximal value of   $\tilde{a}_0$.
Only particles located initially within few $r_0$ experience  maximal wave's intensity.

Qualitatively, in the region $r \geq r_0$ (recall that $r_0 \sim 10^2 R_{NS}$).
\ba &&
\tilde{a}_0 \approx \left( \frac{r}{r_0} \right)^2
\nn && 
\gamma_{\parallel} \approx \frac{\tilde{a}_0}{2}
\ea
For a pulse with rise time $\tau \sim R_{NS}/c$ the distance  $r_{over} $ the pulse would overtake the particle estimates to 
\be
r_{over} \sim c \tau \gamma_{\parallel}^2 
\ee
The condition $ r_{over} \sim c/\Omega$ then gives
\ba && 
\gamma_{\parallel} \sim \tilde{a}_0^{(eff)}/ 2 = (\tau \Omega)^{-1/2} = 70 \sqrt{P}
\ea
where period $P$ is in seconds. The parameter 
\be
 \tilde{a}_0^{(eff)} \sim 10^2 \sqrt{P}
 \ee
  is  a typical nonlinearity parameter that a particle experiences while surfing the pulse. It is  an order of magnitude smaller than would be inferred  without ponderomotive acceleration, Eq. {\ref{aomax}).

\subsection{Ponderomotive and adiabatic acceleration} 
The above results, integration of the parallel momentum  (\ref{betaz}),   did not take adiabatic acceleration into account.  
To further clarify the situation, in Fig. \ref{delay-f1} numerically integrate particles motion in the field of incoming pulse, in inhomogenous decreasing guide field.  This is done using in-house built  Boris-based  pusher \citep{boris_69,birdsall}.  In the simulations a particle  is initially  located 4 rump-up scales ahead of the  $  \tanh (r-t)/\tau$ with rump-up scale $\tau = 5 $ wavelength. At the initial  location of the particle the wave intensity correspond to $\tilde {a}_0=100$. Four different parameters $f$ are shown: $f=100, 10,1$. These different values of $f$ mimic different initial locations in the \ms:  $f=100$ corresponds to $r=r_0$, $f=1$ corresponds to $r=r_f$. 
 We plot the  delay between a local position of a particle and a middle of the pulse, where intensity is half the local maximum. 
 
 Our numerical results indicate that  combined effects of ponderomotive and adiabatic acceleration hugely  increase over-take time.
 We observe that for $f=100$ (lower curve, equivalent to starting at $ r=r_0$) the head of the  pulse quickly passes the particle (since initially its velocity is only mildly relativistic). As a result a particle quickly ``feels'' the full intensity of the wave. In contrast, the head of the pulse never overtakes   a particle located midway between $r_0$ and $r_f$ (middle curve), or further out. This demonstrates that due to a limited radial extent a particle may never reach the terminal  state with local $p_0 = \tilde{a}_0$, as predicted by (\ref{p0ofa0}).
 
 \begin{figure}[h!]
\includegraphics[width=.99\linewidth]{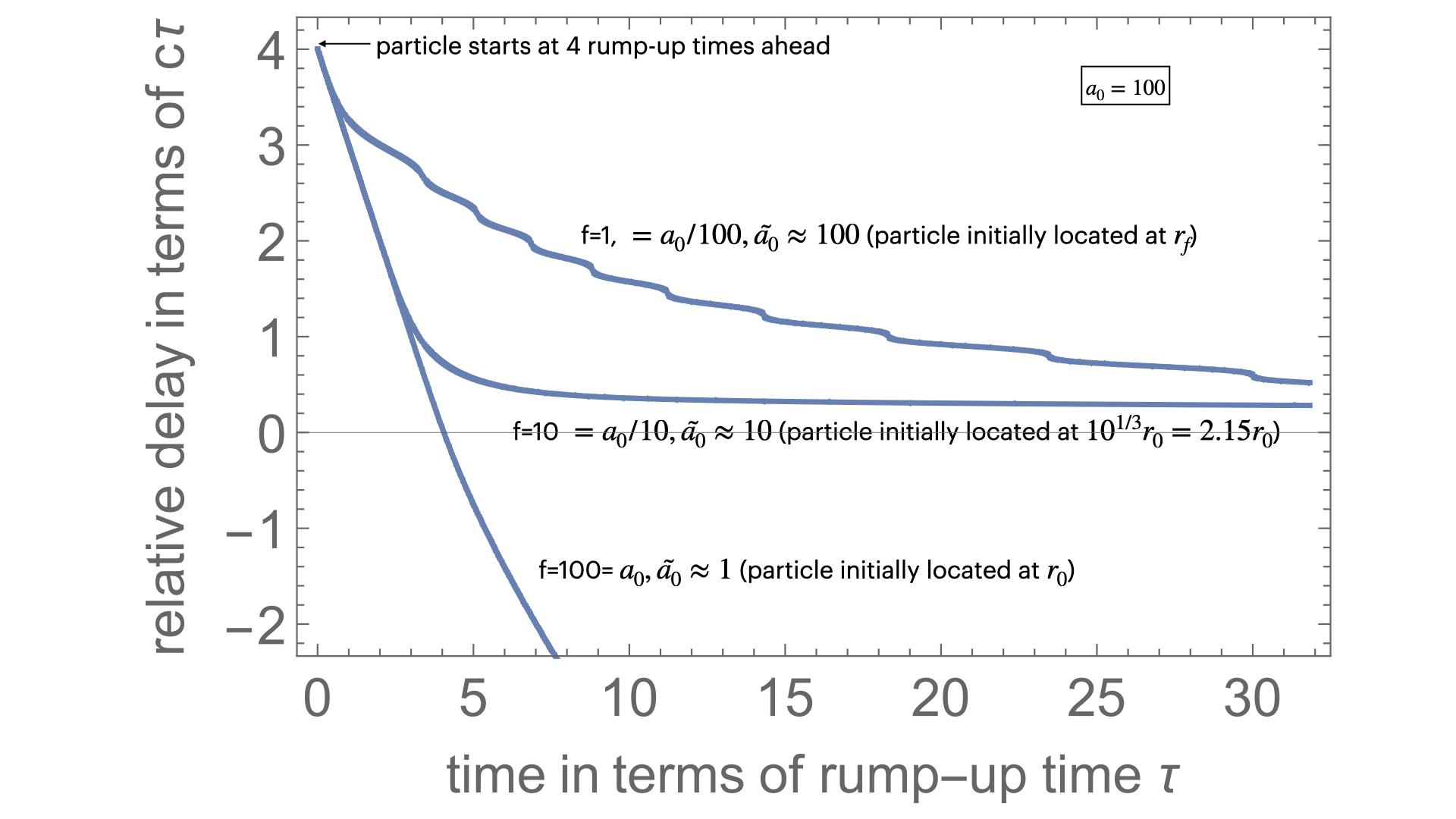}
\caption{ Delay times of ``particle-center of the pulse'' for different initial $f= \om_B/\om$ parameters (effective different initial locations). Calculations start at four rump-up times,  $\tanh$ profile the pulse, \Bf\ changes on 10   ramp-up scale.  }
\label{delay-f1}
\end{figure}

In all cases particles initially located beyond $r_0$ quickly acquire relativistic velocities. For particles near $r_0$ these relativistic velocities are not sufficient to escape  the full nonlinearity of the  wave, bit still, $v\sim c$. Thus, the  leading part of the wave will clear particles from the \ms

\subsection{Dissipated energy}

Let us estimate the  expected energy loss by the wave, assuming that the frontal  particles loses all the energy it acquired in the wave
As figure (\ref{tildeaatpart1}) indicates, only particle in  a narrow layer near $r_0$ (where $\tilde{a}_0$ just becomes of the order of unity) experience large acceleration. Most of the \mss\ surfs the wave and do reach  very large energies. As an estimate of the dissipated power we can use
\begin{itemize}
\item volume $ V  \sim 4\pi r_0^2 \times 3 r_0$ (assuming thickness of $3 r_0$).
\item \Lf\  $\gamma \sim  \left( \tilde{a}_0^{(max)}  \right)^2 /2$
\item density $\kappa n_{GJ}$ ($\kappa \sim 10^5$ is multiplicity, $n_{GJ}$ is the \cite{1970ApJ...160..971G} density).
\item period $P$ in seconds 
\end{itemize}
We find for dissipated energy $E_{dis} $
\be
E_{dis} =5 \times 10^{36} \left( \frac{  \kappa}{10^5} \right)  P^{-1}\, {\rm erg}
\ee
A safely mild value, much smaller than the total energy of the FRB (\ref{Eiso}).

 Fig. \ref{aoefftanh} also indicates that the bulk of the particles in the \ms\  experience $\tilde{a}_0 \sim 100$ (hence $\gamma \sim 5 \times 10^3$). The total associated energy within the \LC\ is then 
 $\sim  5 \times 10^{33}$ ergs. Thus, most of the energy the wave spends on cleaning the \ms near $r\sim r_0$. 

Finally, in Fig. \ref{tildeaatpart1} we plot the non-linearity parameter $\tilde{a}_0$  at the location of a particle in dipolar-like $B \propto r^{-3}$  and monopolar-like $B \propto r^{-2}$ scaling of the guide field.
The rise time of the pulse is assumed to be very short $R_{NS}/c= 30$ micro-seconds. 
Maximum value of the nonlinearity parameter at the location of the particle can reach $\sim 10^3$, but only for sufficiently slow spins $P\geq 100$ msec. If the \ms\ is modified by the ejected CME to have a monopolar-like structure, the maximal non-linearity parameter that a particle experiences is only few times  $10^2$.  Longer rump-up times will further stretch the curves

 \begin{figure}[h!]
\includegraphics[width=.99\linewidth]{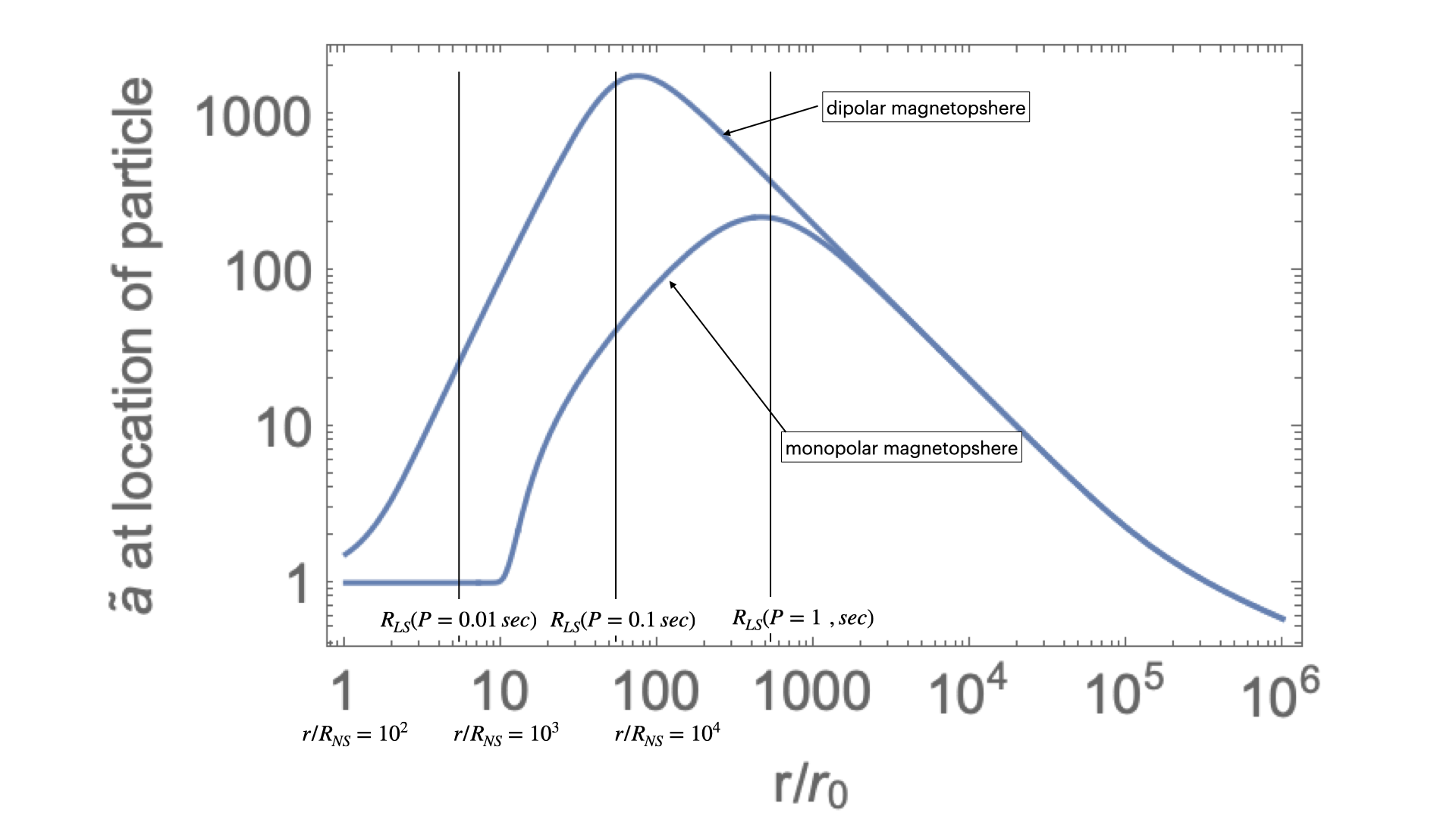}
\caption{Non-linearity parameter $\tilde{a}_0$  at the location of a particle in dipolar-like  and mono-polar-like scaling of the guide field.  A particle is tracked from the point  $r=r_0 = 10^2 R_{NS}$, Eq. (\ref{r0}), where $\tilde{a}_0$ first becomes unity.  Vertical lines indicate location of the \LC\ for different periods $P=0.01, \, 0.1,\, 1 $ second. The assumed rise-time of the FRB pulse is $R_{NS}/c=30$ microseconds.}
\label{tildeaatpart1}
\end{figure}

\section{Different polarization and obliquity}

We also  considered effects of wave's oblique propagation with respect to \Bf, as well as various   polarization.
 Taking  care of the ponderomotive acceleration of particles as the FRBs' wave comes into plasma is most important. The code reproduces well analytical results for ponderomotive acceleration, Fig. \ref{pzoff}.  We run a few simulations for different angles of propagation with respect to the \Bf, and different polarizations.

Besides the parameters $a_0,\, \delta$ and $f$ defined above,  there is angle $\theta$ between the direction of wave propagation and the external \Bf\ and the  polarization angle of the wave $\phi$ (for $\phi =0 $ the \Ef\ of the wave is in the plane defined by the direction of wave propagation and the external \Bf, this is then  the O-mode, for $\phi =\pi/2$ the \Bf\  of the wave is in the plane defined by the direction of wave propagation and the external \Bf, this is then the X-mode). 
 In this particular section, the wave intensity is also modulated by a Gaussian envelope (adiabatic switching  on and off), Fig \ref{gammaoft}.

We observe two types of wave-particle interaction occur. First, for exactly parallel propagation,  as a wave packet come in, it accelerate a particle along the external \Bf\ by the ponderomotive force, plus a particle oscillates in the combined field of the wave and the external \Bf. This motion is reversible:  a particle comes back to rest after the wave have left (if radiation reaction is neglected). 
Second, for oblique propagation,  in a sufficiently strong wave a particle may  occasionally experience DC-type acceleration \citep{2021ApJ...922L...7B}. The acceleration is of the diffusive type:  occasionally a particle  efficiently surfs  the wave gaining energy. Appearance of regions with $E \geq B$ greatly helps this type of acceleration, but is not needed: the O-mode, where $B$ is always larger than $E$ also shows this type of acceleration. This second type of wave-particle energy exchange it dissipative: after the wave has left,  the particle retains some energy. Thus this reduces the wave intensity. 

We find that the dissipative interaction is highly dependent on the obliquity and polarization.
The particular case considered by  \cite{2021ApJ...922L...7B},  that of an X-mode propagating perpendicular to the \Bf\ is the most extreme, the most dissipative one.  For more general geometry and polarization, the resulting energy  exchange is orders-of-magnitude smaller. Any pre-wave parallel motion further reduces the losses.

\begin{figure}[h!]
\includegraphics[width=0.99\linewidth]{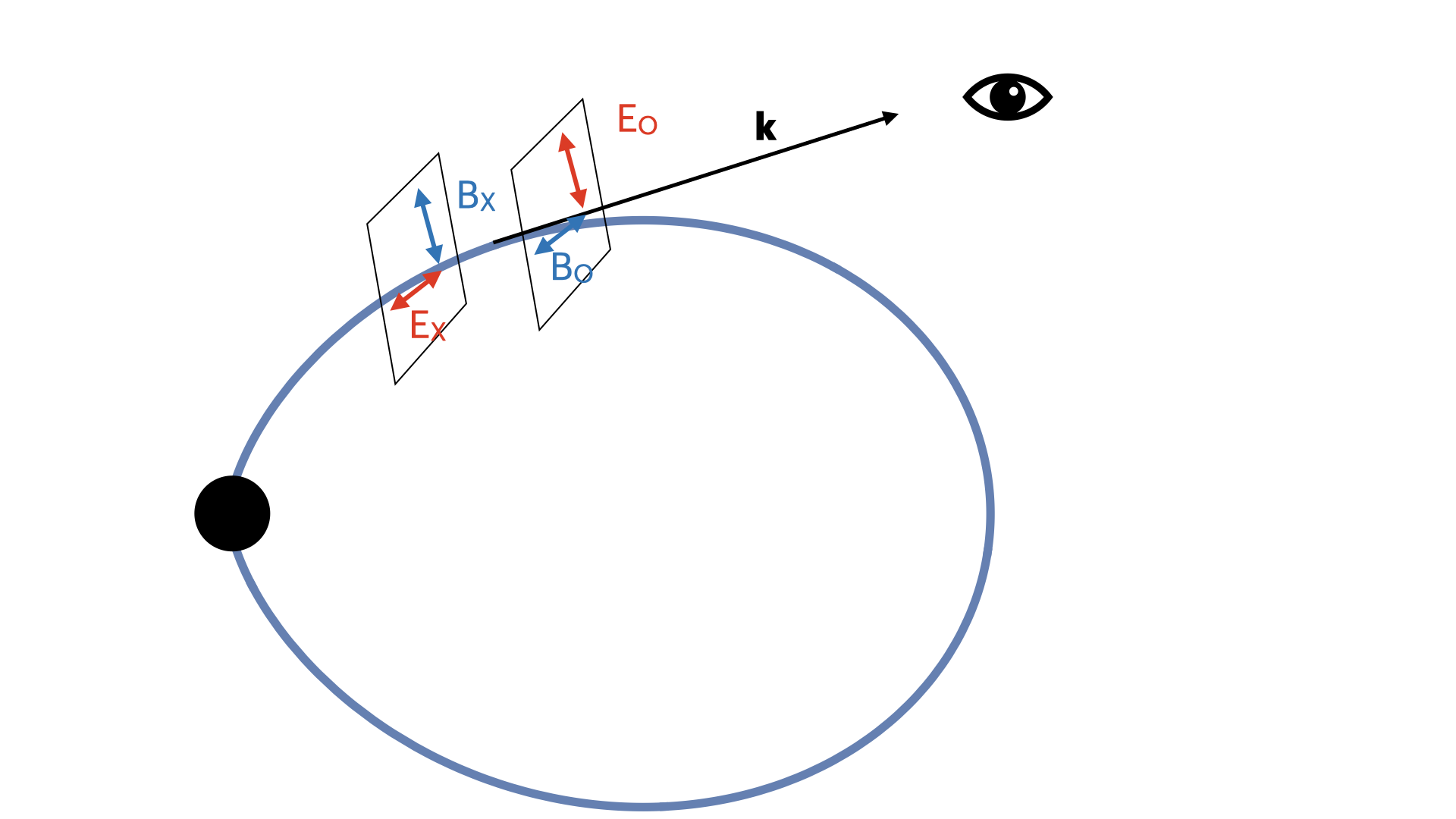}
\caption{Two polarizations add differently with the external \Bf. For the X-mode  the \Bf\ of the wave is in the plane of the dipolar field; the total \Bf\ at some moments may become smaller than the \Ef.  For the O-mode  the \Bf\ of the wave is perpendicular to the dipolar field; the total \Bf\  is always larger  than the \Ef. }
\label{01}
\end {figure}

\begin{figure}[h!]
\includegraphics[width=0.49\linewidth]{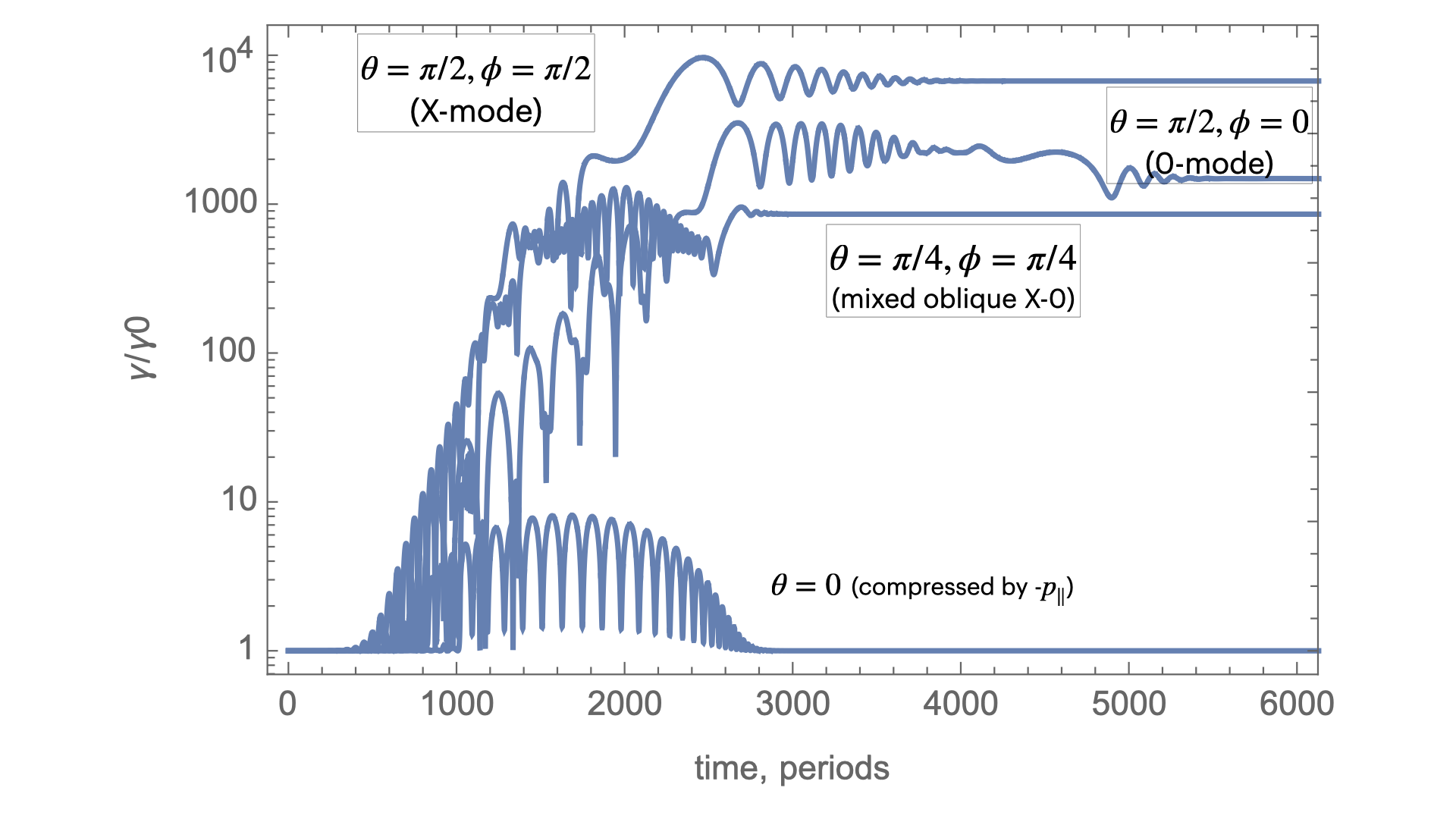}
\i\includegraphics[width=0.49\linewidth]{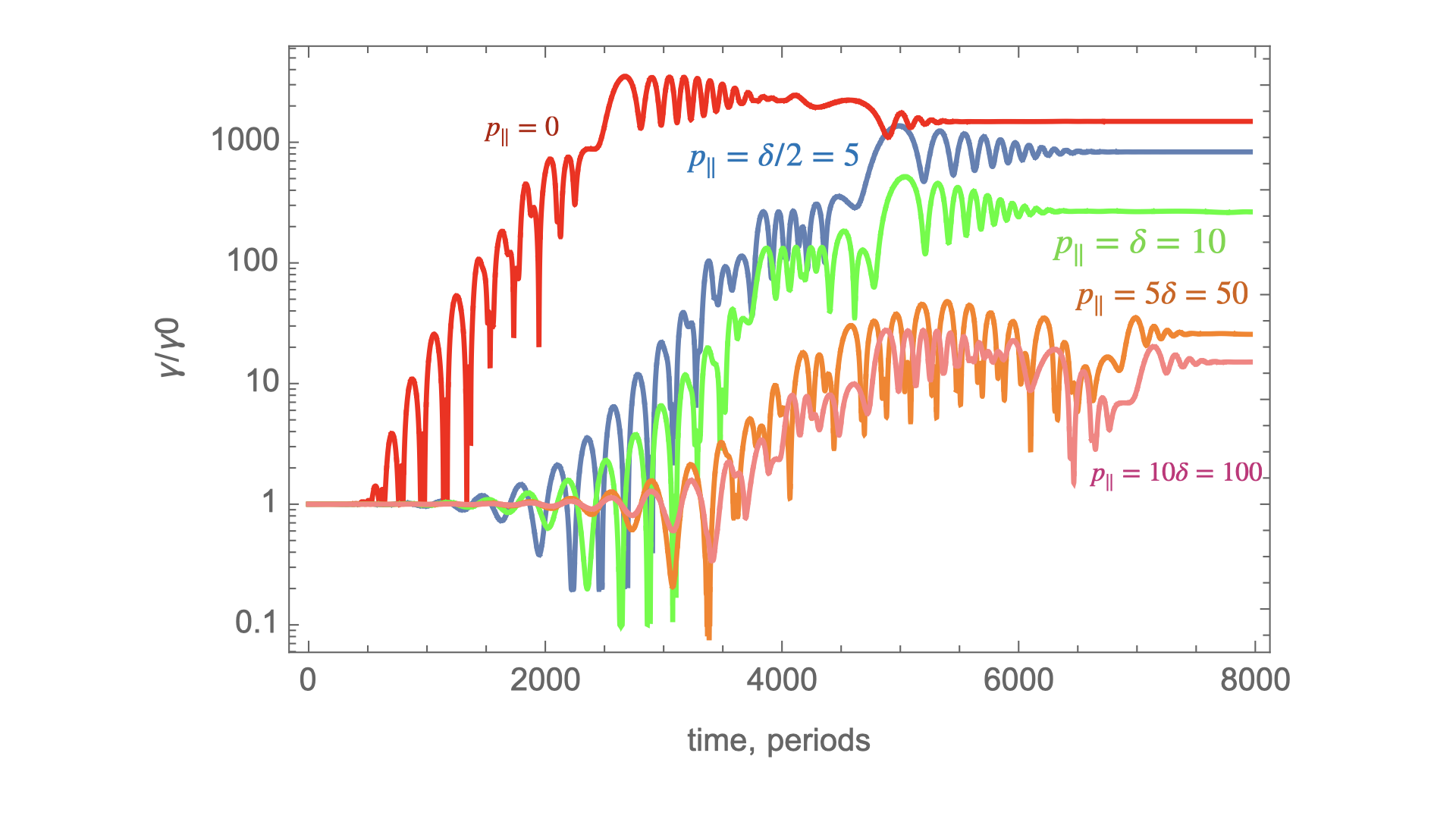}
\caption{Left panel:  Evolution of \Lf\ for a laser pulse with  $\delta=10$, $f=100$ for different directions of propagation with respect to the \Bf\ (angle $\theta$)  and  different polarization (angle $\phi$).In the parallel case  $\theta=0$ the ponderomotive force accelerates the particle along \Bf\ so that it surfs the wave  for  a very long time; to shorten  dynamical time the particle was injected with momentum against the wave. Most importantly, in the parallel case there is no dissipation: after the wave packet passes by,  the particle energy returns to the initial \Lf. 
Both for X-mode and O-mode periods of just oscillations in the field are intermitted with ``surfing'' - a sudden  increase of particle's energy. 
 The X-mode is most prone to dissipation/particle acceleration. The O-mode  experiences an order-of-magnitude smaller dissipation. Parallel propagation is completely ideal.
 Right panel: Influence of the initial outward momentum. Evolution of the {\it relative}  \Lf\ of particle moving initially along \Bf\  with various {\Lf}s. Parameters: $\delta =10$, $f=100$, $\theta = \pi/4$, $\phi = \pi/4$.
}
\label{gammaoft}
\end {figure}

Our numerical simulations imply
\begin{itemize}
\item X-mode propagating perpendicular to the \Bf\ indeed suffers heavy absorption. This is the case considered by \cite{2021ApJ...922L...7B}. In the X-mode, the guide  \Bf\   periodically subtracts from the  wave's \Bf\, leading to appearance $E\geq B_{tot} $ and efficient particle acceleration
\item O-mode, for which $E \leq B$ always, also shows occasional burst of particle's acceleration. In these cases the particles {\it nearly} surfs the \Ef\ of the wave. But overall, the energy gained by a particle in the O-mode is an order of magnitude smaller that in the X-mode.
\item Exactly parallel propagation is purely non-dissipative
\item initial parallel motion away from the pulse greatly reduce dissipative effects.
\end{itemize}

\section{Effects of initial parallel velocity}

Opening of the \ms, \S \ref{Post-eruption}, will also generate radial plasma outlfow.  We did a series   of  numerical  runs that  included initial parallel motion of a particle,  Fig \ref{gammaoft}.   
Our conclusion is that initial parallel momentum greatly deacreses the efficiency of wave-particle interaction.

  To estimate the effects of initial parallel momentum, we note that instead of (\ref{para11}) we have
  \ba  &&
  \gamma-p_z = \gamma_i - p_i
  \nn && 
   \gamma_i = \sqrt{1+p_i^2}
   \ea
   where $\gamma_i$ and $  p_i$ are  initial \Lf\ and mmentum along the field (away from the diretion of pulse propagation). 
   
   We find
   \ba &&
   p_z = \frac{1+p_0^2/2} {\gamma_i - p_i} - \gamma_i \approx (1+p_0^2) \gamma_i
   \nn &&
   \gamma = \sqrt{ 1+p_0^2 + \frac{  \left(1+ p_0^2/2 -\gamma _i^2+\gamma _i p_i\right){}^2}{\left(\gamma _i- p_i\right){}^2} } \approx  (1+p_0^2) \gamma_i
   \nn &&
   \beta_z = p_z/\gamma\approx 1- \frac{1}{2 (1+p_0^2) \gamma_i^2}
   \nn &&
   \gamma_\parallel \approx \sqrt{ (1+p_0^2) }\gamma_i
   \ea
   
   The resonant condition (\ref{omres}) now gives, approximately, $\gamma_ i \gg1$,
   \be
   p_0 = \frac{a_0}{1+ \gamma_i^{3/2} /(2 \sqrt{2}) f} \approx \gamma_i^{-3/2}  \frac{a_0}{f},
   \ee
   showing that even mild values of $ \gamma_i \sim $ few strongly suppress wave-particle interaction.
     Qualitatively, initial parallel motion with \Lf\ $\gamma_i$ away from the star reduces the inital wave's frequency in the particle frame, leading to higher effective $f$, and decrease of $\tilde{a}_0$.

\section{Self-cleaning}
\label{Self-cleaning}

As discussed above the most important dissipation occurs on particles that start from $r\sim r_0 \sim 10^2 R_{NS} $ (where $\tilde{a}_0 \sim 1$), and experience largest dissipation at $r \sim r_f \sim 5\times 10^3 R_{NS}$ where  $f \sim 1$. As the  ratio $r_f /r_0 \gg 1$, the field geometry (with respect to wave's propagating) may change substantially.  

For mildly oblique propagation, $\theta \neq 0$, the surfing effect is reduced. For small angles of propagation   a particle still  surfs the wave for a long time $\sim 1/\theta$. At large angles, $\theta \sim 1$, a new effect appears - self cleaning, Fig. \ref{Self-cleaning-fig}.  Consider a pulse of finite transverse dimensions. For oblique \Bf\ the leading part of the pulse would ponderomotively push the plasma particles along the local \Bf. For oblique \Bf, particles will stream sideways, clearing the path for the main part of the pulse.

Fig. \ref{Self-cleaning-fig}, right panel,  shows the results of simulations. Plotted is a transverse (sideway) displacement of a particle as a function of magnetic obliquity $\theta$ for several parameters of  wave intensity $\delta$ and the frequency parameter $f$. Overall, the displacement shows the expected $\propto \sin \theta \cos \theta$ dependence.  Except for the case of mild wave in high \Bf\  ($\delta  =1,\, f=10$), the curves approximately match, since  the motion along the field becomes relativistic.  

In addition, intermediate cases  show relatively high random  variations in the final displacement. This generally consistent with the previously discusses concept that particles'  trajectories can be mildly stochastic. 

\begin{figure}[h!]
\includegraphics[width=0.49\linewidth]{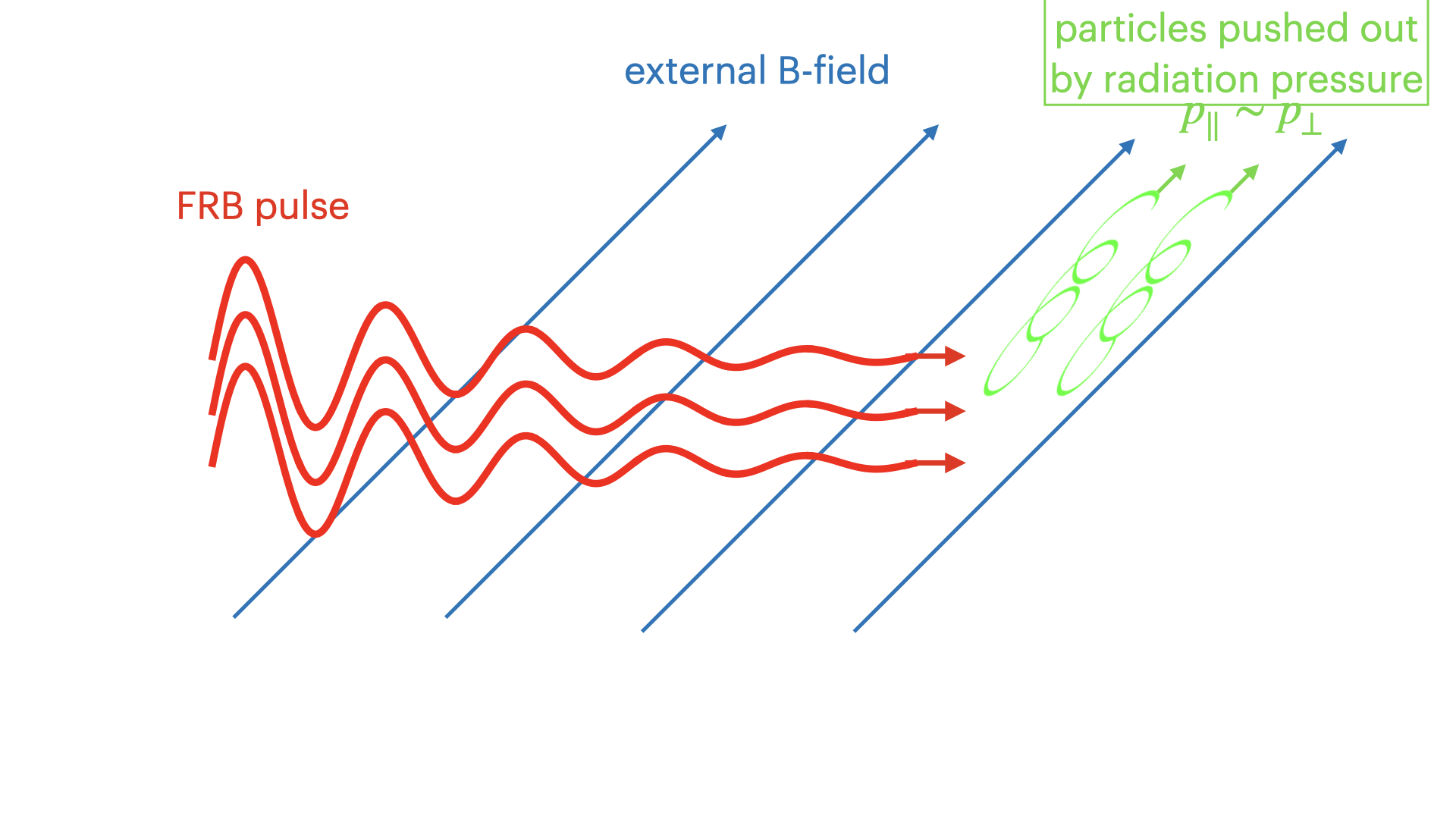}
\includegraphics[width=0.49\linewidth]{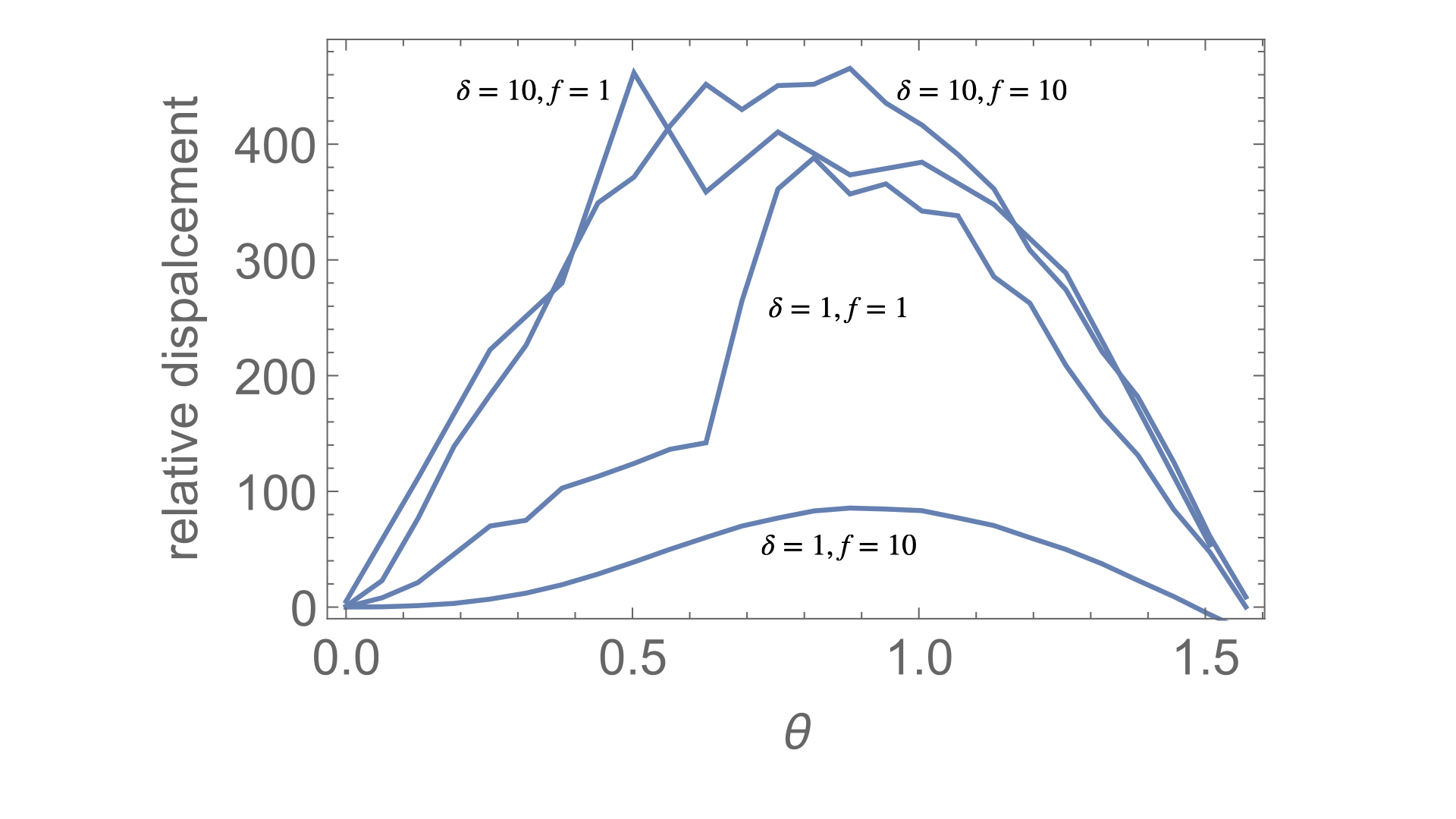}
\caption{Self-cleaning. Left panel: cartoon how the leading part of the pulse, propagating at oblique angles to the \Bf, ponderomotively pushes the particles away from the propagation path. The bulk of the pulse then propagates through nearly vacuum, and does not experience dissipation. Right panel: sideway displacement as function of the angle between the waves' propagation and \Bf.}
\label{Self-cleaning-fig}
\end {figure}

\section{Discussion}

In this work  we consider escape of high brightness radiation  from magnetars' \mss, and conclude that there are multiple ways to avoid nonlinear absorption. 
First, strong non-linear effects are expected in a limited range of parameters, around surface \Bfs\ of $\sim 10\% $ of critical and spin periods of $\sim$ tens of millisecond. Larger \Bfs, and shorter periods limit the effective non-linearity parameter to $\tilde{a}_0 \leq 10^2$, see Fig. \ref{aomaxofP1}.

In the ``region of interest", we find that ponderomotive and parallel-adiabatic acceleration of particles.
are  most important.  In a mildly strong leading part of the wave,  particles quickly get large parallel momenta - this effectively freezes the interaction. Roughly speaking, in order to obtain transverse  momentum $\sim p_0$, a particle needs to serf for time $\sim p_0^2 \tau$, where $\tau$, is a rump-up time of the \EM\ pulse. In the inner parts of the \ms, the value of $p_0$ is suppressed by the guide \Bf, so a pulse passes through quickly, bit does not  shake the particles much. Further out, where a stationary particle  could have  been   accelerated  to a large \Lf, it never happens because a particle is surfing the pulse and remains at locally low non-linearity parameter, before  escaping \ms.

All  these issues are further  overwhelmed by the  parallel large \Lf\ along the newly  opened \Bf\ lines, possibly  initiated by the opening of the \ms\  during a CME. Large parallel momentum reduces $\delta = B_w/B_0$ in the particles's frame, and leads to further freezing of the wave-particle dynamics. 


Other  points are:  (i)  the O-mode (which never has regions  $E\geq B$ is much less prone to dissipation  (but particles in the wave  still  experience  occasional energy boost, draining wave's energy);  (ii) quasi-parallel propagation is intrinsically non-dissipative; (iii) magnetospheric dynamics during magnetar's explosions ensures that the \Bf\ becomes nearly  radial beyond some distance  - smaller than the \LC\ and dependent on the power of the explosion,  see Eq. (\ref{req});  as a result the \EM\ pulse propagates nearly along the \Bf; (iv) initial mildly relativistic velocity along the field, away  from the star further reduces particles' losses; (iv) leading part of the pulse may push the plasma sideways, clearing the path for the main  part of the pulse.

We conclude that the case considered by \cite{2021ApJ...922L...7B,2022PhRvL.128y5003B,2023arXiv230712182B}, X-mode propagating equatorially across \Bf,   is extreme, and is not indicative of the general situation. 
That is a specific case of no-surfing.

In the approach of     \cite{2023arXiv230909218G}, the  energy density of the waves and their frequency (the transformation rate)   should be calculated in plasma frame, which is flying away with large Lorentz factor. In that frame  the wave's energy density   is down by induced parallel \Lf\ $ \gamma_\parallel^2$ and frequency is  down by $\gamma_\parallel$ , so  total reduction of the efficiency of nonlinear interaction is $\sim \gamma_\parallel^3$.

 This work had been supported by 
NASA grants 80NSSC17K0757 and 80NSSC20K0910,   NSF grants 1903332 and  1908590.
I would like to thank  Alexey Arefiev, Andrei Beloborodov, Pawan Kumar,  Mikhail Medvedev, Kavin Tangtartharakul, Chris Thompson,  and  Bing Zhang for discussions. 

\section{Data availability}
The data underlying this article will be shared on reasonable request to the corresponding author.

\bibliographystyle{apj}

 \bibliography{/Users/maxim/Dropbox/Research/BibTex} 

\end{document}